\documentclass[12pt]{article}
\usepackage[latin9]{inputenc}
\usepackage{geometry}
\usepackage{amsmath}
\usepackage{amssymb}
\usepackage{stmaryrd}
\usepackage{wasysym}
\usepackage[authoryear]{natbib}

\makeatletter
\usepackage{geometry}
\usepackage{amsmath}
\usepackage{amssymb}
\usepackage{amsfonts}
\usepackage{amsmath, amssymb,bm, amsthm, stmaryrd}
\usepackage{lscape}
\usepackage{appendix}
\usepackage{graphicx}
\usepackage{setspace}
\usepackage{dcolumn}
\usepackage[authoryear]{natbib}
\usepackage{rotating}
\usepackage{lscape}
\usepackage{comment}
\usepackage{color}
\usepackage{breakurl}
\usepackage[unicode=true,pdfusetitle,
 bookmarks=true,bookmarksnumbered=false,bookmarksopen=false,
 breaklinks=false,pdfborder={0 0 1},backref=section,colorlinks=false]
 {hyperref}
\hypersetup{
 colorlinks,citecolor=blue,pdftex}

\usepackage{tikz}
\usepackage{caption}
\usepackage{subcaption}
\usepackage{tkz-tab}
\usepackage{tikz-3dplot}
\usetikzlibrary{calc}  

\newcolumntype{d}[1]{D{.}{.}{#1}}
\newcolumntype{t}[1]{D{,}{,}{#1}}
\newcolumntype{i}[1]{D{.}{}{#1}}
\newtheorem{theorem}{Theorem}[section]

\newtheorem{corollary}{Corollary}[section]

\newtheorem{lemma}{Lemma}[section]

\newtheorem{remark}{Remark}[section]

\theoremstyle{plain} 

\newtheorem*{asC}{Assumption C}
\newtheorem*{asSX}{Assumption SX}
\newtheorem*{asRS}{Assumption RS}
\newtheorem*{asSP}{Assumption SP}

\newtheorem*{asC2}{Assumption C$^{\prime}$}
\newtheorem*{asSX2}{Assumption SX$^{\prime}$}
\newtheorem*{asRS2}{Assumption RS$^{\prime}$}
\newtheorem*{asSP2}{Assumption SP$^{\prime}$}

\numberwithin{equation}{section}
\makeatother

\begin{document}

\title{Identification in Nonparametric Models \\
for Dynamic Treatment Effects\thanks{The author is grateful to Dan Ackerberg, Tim Armstrong, Stephane Bonhomme,
Xiaohong Chen, Jim Heckman, Pedro Sant'Anna, Ed Vytlacil and Nese
Yildiz for their helpful discussions. Also, comments from participants
in seminars at UChicago, Rochester, USC, Penn State, Indiana, NUS,
SNU, and in the 2018 Asian Meeting and China Meeting of the Econometric
Society, and the 2018 International Panel Data Conference are appreciated.}}

\author{Sukjin Han\\
 Department of Economics\\
 University of Texas at Austin \\
\href{mailto:sukjin.han@austin.utexas.edu}{sukjin.han@austin.utexas.edu}}

\date{First Draft: August 12, 2017 \\
 This Draft: \today}

\maketitle
\bigskip{}

\begin{abstract}
This paper develops a nonparametric model that represents how sequences
of outcomes and treatment choices influence one another in a dynamic
manner. In this setting, we are interested in identifying the average
outcome for individuals in each period, had a particular treatment
sequence been assigned. The identification of this quantity allows
us to identify the average treatment effects (ATE's) and the ATE's
on transitions, as well as the optimal treatment regimes, namely,
the regimes that maximize the (weighted) sum of the average potential
outcomes, possibly less the cost of the treatments. The main contribution
of this paper is to relax the sequential randomization assumption
widely used in the biostatistics literature by introducing a flexible
choice-theoretic framework for a sequence of endogenous treatments.
This framework allows non-compliance of subjects in experimental studies
or endogenous treatment decisions in observational settings. We show
that the parameters of interest are identified under each period's
two-way exclusion restriction, i.e., with instruments excluded from
the outcome-determining process and other exogenous variables excluded
from the treatment-selection process. We also consider partial identification
in the case where the latter variables are not available. Lastly,
we extend our results to a setting where treatments do not appear
in every period.

\vspace{0.1in}


\end{abstract}

\section{Introduction\label{sec:Introduction}}

This paper develops a nonparametric model that represents how sequences
of outcomes and treatment choices influence one another in a dynamic
manner. Often, treatments are chosen multiple times over a horizon,
affecting a series of outcomes. Examples are medical interventions
that affect health outcomes, educational interventions that affect
academic achievements, job training programs that affect employment
status, or online advertisements that affect consumers\textquoteright{}
preferences or purchase decisions. Agents endogenously make decisions
of receiving treatments, e.g., whether to comply with random assignments.
The relationship of interest is dynamic in the sense that the current
outcome is determined by past outcomes as well as current and past
treatments, and the current treatment is determined by past outcomes
as well as past treatments. Such dynamic relationships are clearly
present in the aforementioned examples.  A static model misrepresents
the nature of the problem (e.g., nonstationarity, state dependence,
learning) and fails to capture important policy questions (e.g.,
optimal timing and schedule of interventions).

In this setting, we are interested in identifying the dynamic causal
effect of a sequence of treatments on a sequence of outcomes or on
a terminal outcome that may or may not be of the same kind as the
intermediate outcomes. We are interested in learning about the average
of the outcome in each period, had a particular treatment sequence
been assigned \textit{up to that period}, which defines the potential
outcome in this dynamic setting. We are also interested in the average
treatment effects (ATE's) and the transition-specific ATE's defined
based on the average potential outcome, unconditional and conditional
on the previous outcomes, respectively. For example, one may be interested
in whether the success rate of a particular outcome (or the transition
probability) is larger with a sequence of treatments assigned in relatively
later periods rather than earlier, or with a sequence of alternating
treatments rather than consistent treatments. The treatment effect
is said to be dynamic, partly because the effect can vary depending
upon the period of measurement, even if the same set of treatments
is assigned. Lastly, we are interested in the optimal treatment regimes,
namely, sequences of treatments that maximize the (weighted) sum of
the average potential outcomes, possibly less the cost of the treatments.
For example, a firm may be interested in the optimal timing of advertisements
that maximizes its aggregate sales probabilities over time, or a sequence
of educational programs may be aimed to maximize the college attendance
rate. We show that the optimal regime is a natural extension of a
static object commonly sought in the literature, namely, the sign
of the ATE. Analogous to the static environment, knowledge about
the optimal treatment regime may have useful policy implications.
For example, a social planner may wish to at least exclude specific
sequences of treatments that are on average suboptimal.

Dynamic treatment effects have been extensively studied in the biostatistics
literature for decades under the counterfactual framework with a sequence
of treatments (\citet{robins1986new,robins1987graphical,robins1997causal},
\citet{murphy2001marginal}, \citet{murphy2003optimal}, among others).
In this literature, the crucial condition used to identify the average
potential outcome is a dynamic version of a random assignment assumption,
called the\textit{ sequential randomization}. This condition assumes
that the treatment is randomized in every period within those individuals
who have the same history of outcomes and treatments.\footnote{This assumption is also called sequential conditional independence
or sequential ignorability. In the econometrics literature, \citet{vikstrom2018bounds}
consider treatment effects on a transition to a destination state,
and carefully analyze what the sequential randomization assumption
can identify in the presence of dynamic selection.} This assumption is only suitable in experimental studies with the
perfect compliance of subjects, which is often infeasible (\citet{robins1994correcting,robins2004estimation}).
When interventions continue for multiple periods as in the examples
described above, non-compliance may become more prevalent than in
one-time experiments, e.g., due to the cost of enforcement or the
subjects' learning. In addition to partial compliance in experimental
settings, sequential randomization is invalid in many observational
contexts as well.

The main contribution of this paper is to relax the assumption of
sequential randomization widely used in the literature by establishing
a flexible choice-theoretic framework for a sequence of endogenous
treatments. To this end, we consider a simple nonparametric structural
model for a dynamic endogenous selection process and dynamic outcome
formation. In this model, individuals are allowed not to fully comply
with each period's assignment in experimental settings, or are allowed
to make an endogenous choice in each period as in observational settings.
The heterogeneity in each period's potential outcome is given by recursively
applying a switching-regression type of models with a sequential version
of rank similarity. The joint distribution of the full history of
unobservable variables in the outcome and treatment equations is still
flexible, allowing for arbitrary forms of treatment endogeneity as
well as serial correlation. Relative to the counterfactual framework,
the dynamic mechanism is clearly formulated using this structural
model, which in turn facilitates our identification analysis.

We show that the average potential outcome, or equivalently, the average
recursive structural function (ARSF) given the structural model we
introduce, is identified under a two-way exclusion restriction. That
is, we assume there exist (possibly binary) instruments excluded from
the outcome-determining process and exogenous variables excluded from
the treatment-selection process. A leading example of the former is
a sequence of randomized treatment assignments or randomized encouragements
(\citet{sexton1984clinical}) from, e.g., clinical trials, field experiments,
and A/B testings, and other examples include sequential policy shocks.
Examples of the latter include factors that agents cannot fully anticipate
when making treatment or compliance choices but that determine the
outcome. We show that such timing can be justified in this dynamic
context, and some covariates in the outcome process may be valid candidates.
Identification in nonseparable triangular models using this exclusion
restriction is pioneered by \citet{VY07} and subsequently appears
in \citet{SV11} and \citet{Han2018} among others, all in static
settings. The dynamic structure introduced in this paper poses added
challenges in using a similar strategy, since (i) the outcome and
treatment structural functions depend on the vectors of lags, which
in turn make each potential outcome a direct function of all the previous
potential outcomes, (ii) the period specific knowledge analogous to
that in \citet{VY07} does not directly recover any meaningful objects
of interest in general, (iii) rank invariance substantially restricts
heterogeneity in this dynamic setting, and (iv) the initial condition
problem is present. In this paper, we address these challenges and
show how to achieve identification. In particular, we introduce sets
of unobservable vectors across periods as a simple way to express
potential outcomes in the presence of complicated dynamics. We then
recover period specific knowledge using the exclusion restriction,
which is then iteratively incorporated across periods for identification
by means of mathematical induction, obeying the recursive structure
of the potential outcome. In doing so, we introduce sequential rank
similarity which substantially weakens the naive rank similarity or
rank invariance. The proof is constructive and provides a closed form
expression for the ARSF. The identification of each period's ARSF
allows us to point identify the ATE's and the optimal treatment regimes.
In this paper, we also consider cases where the two-way exclusion
restriction is violated in the sense that only a standard exclusion
restriction holds or that the variation of the exogenous variables
is limited. In these cases, we can calculate the bounds on the parameters.
As an extension of our results, we consider another empirically relevant
situation where treatments do not appear in every period, while outcomes
are constantly observed. We show that the parameters of interest and
the identification analysis can be easily modified to incorporate
this situation.

This paper contributes to growing research on the identification of
the effects of dynamic endogenous treatments that allows for treatment
heterogeneity. \citet{cunha2007identification} and \citet{heckman2007dynamic}
consider a semiparametric discrete-time duration model for the choice
of the treatment timing and associated outcomes. Building on these
works, \citet{heckman2016dynamic} consider not only ordered choice
models but also unordered choice models for up-or-out treatment choices.\footnote{As related works, the settings of \citet{angrist1995two}, \citet{jun2016multiple},
and \citet{lee2016identifying} for multiple (or multi-valued) treatment
effects may be applied to a dynamic setting. Also, see \citet{abbring2007econometric}
for a survey on dynamic treatment effects.} An interesting feature of their results is that dynamic treatment
effects are decomposed into direct effects and continuation values.
As an important feature, these papers consider attrition based on
the irreversible treatment decisions; see also \citet{sasaki2015heterogeneity}.
Similar to our approach, \citet{heckman2007dynamic} and \citet{heckman2016dynamic}
utilize exclusion restrictions. Unlike these papers, however, we do
not necessarily invoke infinite supports of each period's exogenous
variables but instead use the two-way exclusion restriction. \citet{abraham2018estimating},
\citet{athey2018design}, and \citet{callaway2018difference} extend
a difference-in-differences approach to dynamic settings without specifying
fixed-effect panel data models. They consider the effects of treatment
timing on the treated, where the treatment process is irreversible
as in the previous works. Unlike all the papers mentioned in this
paragraph, we consider nonparametric dynamic models for treatment
and outcome processes with a general form of evolution, where the
processes can freely change states. These models can include an irreversible
process as a special case. Moreover, we consider different identifying
assumptions than those in the previous works and focus on the identification
of the ATE's and related parameters.

This paper's structural approach is only relative to the counterfactual
framework of Robins. A fully structural model of dynamic programming
is considered in the seminal work by \citet{rust1987optimal} and
more recently by, e.g., \citet{blevins-2014} and \citet{buchholz2016semiparametric}.
This literature typically considers a single rational agent's optimal
decision, whereas we consider a large group of heterogenous agents
with no assumptions on agents' rationality or strong parametric assumptions.
Most importantly, our focus is on the identification of the effects
of treatments formed as agents' decisions.  The agnostic approach
of this paper is, in spirit, similar to \citet{heckman2007dynamic}
and \citet{heckman2016dynamic}, in that we remain flexible for the
economic and non-economic components of the model. Lastly, \citet{torgovitsky2016partial}
extends the literature on dynamic binary response models (with no
treatment) by considering a counterfactual framework without imposing
parametric assumptions. In his framework, the lagged outcome plays
the role of a treatment for the current outcome, and the ``treatment
effect'' captures the state dependence. Here, we consider the effects
of the treatments on the outcomes, and introduce a selection equation
for each treatment as an important component of the model. As an extension
of our analysis, we identify the transition-specific ATE, which is
related to the effect of a treatment on the state dependence. 

In the next section, we first introduce Robins's counterfactual outcome
framework and discuss sequential randomization. Section \ref{sec:model_and_para}
introduces the main structural model of this paper with parameters
of interest, followed by a motivating example in Section \ref{sec:Motivating-Example}.
The main identifying conditions and identification results are present
in Section \ref{sec:ID}, and several extensions are discussed in
Sections \ref{sec:Treatment-Effects-on}\textendash \ref{sec:subseq}.
Section \ref{sec:Conclusions} briefly concludes. In the Appendix,
all the proofs are collected and estimation and inference are discussed.

In terms of notation, let $\boldsymbol{W}^{t}\equiv(W_{1},..,W_{t})$
denote a row vector that collects r.v.'s $W_{t}$ across time up to
$t$, and let $\boldsymbol{w}^{t}$ be its realization. Note $\boldsymbol{W}^{1}=\boldsymbol{W}_{1}$.
We sometimes write $\boldsymbol{W}\equiv\boldsymbol{W}^{T}$ for convenience.
For a vector $\boldsymbol{W}$ without the $t$-th element, we write
$\boldsymbol{W}_{-t}\equiv(W_{1},...,W_{t-1},W_{t+1},...,W_{T})$
with realization $\boldsymbol{w}_{-t}$. More generally, let $\boldsymbol{W}_{-}$
with realization $\boldsymbol{w}_{-}$ denote some subvector of $\boldsymbol{W}$.
Lastly, for r.v.'s $Y$ and $W$, we sometimes abbreviate $\Pr[Y=y|W=w]$
and $\Pr[Y=y|W\in\mathcal{W}]$ to $\Pr[Y=y|w]$ (or $P[y|w]$) and
$\Pr[Y=y|\mathcal{W}]$, respectively.

\section{Robins's Framework\label{sec:Robins}}

We first introduce Robins's counterfactual framework and state the
assumption of sequential randomization commonly used in the biostatistics
literature (\citet{robins1986new,robins1987graphical}, \citet{murphy2001marginal},
\citet{murphy2003optimal}). For a finite horizon $t=1,...,T$ with
fixed $T$, let $Y_{t}$ be the outcome at $t$ with realization $y_{t}$
and let $D_{t}$ be the binary treatment at $t$ with realization
$d_{t}$. The underlying data structure is panel data with a large
number of cross-sectional observations over a short period of time
(and the cross-sectional index $i$ suppressed throughout, unless
necessary). We call $Y_{T}$ a \textit{terminal outcome} and $Y_{t}$
for $t\le T-1$ a \textit{intermediate outcome}.\footnote{The terminal period $T$ may be an administrative end of follow-up
time.} Let $\mathcal{Y}$ and $\mathcal{D}\subseteq\{0,1\}^{T}$ be the
supports of $\boldsymbol{Y}\equiv(Y_{1},...,Y_{T})$ and $\boldsymbol{D}\equiv(D_{1},...,D_{T})$,
respectively. There can be other time-varying covariates present in
this setup, but we omit them here.

Consider a treatment regime $\boldsymbol{d}\equiv(d_{1},...,d_{T})\in\mathcal{D}$,
which is defined as a predetermined hypothetical sequence of interventions
over time, i.e., a sequence of each period's decisions on whether
to treat or not, or whether to choose treatment $A$ or treatment
$B$.\footnote{This is called a nondynamic regime in the biostatistics literature.
A dynamic regime is a sequence of treatment assignments, each of which
is a predetermined function of past outcomes. A nondynamic regime
can be viewed as its special case, where this function is constant.
See, e.g., \citet{murphy2001marginal,murphy2003optimal} for related
discussions.} Then, a potential outcome at $t$ can be written as $Y_{t}(\boldsymbol{d})$.
This can be understood as an outcome for an individual, had a particular
treatment sequence been assigned. Although the genesis of $Y_{t}(\boldsymbol{d})$
can be very general under this counterfactual framework, the mechanism
under which the sequence of treatments interacts with the sequence
of outcomes is opaque. The definition of $Y_{t}(\boldsymbol{d})$
becomes more transparent later with the structural model introduced
in this paper.

Given these definitions, we state the assumption of sequential randomization
by Robins: For each $\boldsymbol{d}\in\mathcal{D}$,
\begin{align}
(Y_{1}(\boldsymbol{d}),...,Y_{T}(\boldsymbol{d}))\perp D_{t}|\boldsymbol{Y}^{t-1},\boldsymbol{D}^{t-1}\label{eq:seq_rand}
\end{align}
for $t=1,...,T$. This assumption asserts that, holding the history
of outcomes and treatments (and potentially other covariates) fixed,
the current treatment is fully randomized. Sequential randomization
can be violated if agents make decisions $D_{t}$ based on time-varying
or time-invariant factors, unobserved to the analyst. In the next
section, we relax this assumption and specify dynamic selection equations
for a sequence of treatments that are allowed to be endogenous, i.e.,
to be dependent on unobservable factors. Apart from this assumption,
we maintain the same preliminaries introduced in this section.

\begin{remark}[\textbf{Irreversibility}]\label{rem:irrev}As a special
case of our setting, the process of $D_{t}$ may be irreversible in
that the process only moves from an initial state to a destination
state, i.e., the destination state is an absorbing state. The up-or-out
treatment decision (or the treatment timing) can be an example where
the treatment process satisfies $D_{t}=1$ once $D_{t-1}=1$ is reached,
as in \citet{heckman2007dynamic}, \citet{heckman2016dynamic}, \citet{abraham2018estimating}
and \citet{callaway2018difference}. Although it is not the main focus
of this paper, the process of $Y_{t}$ may as well be irreversible.
This case, however, requires caution due to dynamic selection; see
discussions later in this paper. The survival of patients ($Y_{t}=0$)
in discrete time duration models can be an example where the transition
of the outcome satisfies $Y_{t}=1$ once $Y_{t-1}=1$. In this case,
it may be that $D_{t}$ is missing when $Y_{t-1}=1$, which can be
dealt by conventionally assuming $D_{t}=0$ if $Y_{t-1}=1$. When
processes are irreversible, the supports $\mathcal{D}$ and $\mathcal{Y}$
are strict subsets of $\{0,1\}^{T}$.\end{remark}

\begin{remark}[\textbf{Terminal outcome of a different kind}]\label{rem:term}As
in \citet{murphy2001marginal} and \citet{murphy2003optimal}, we
may be interested in a terminal outcome that is of a different kind
than that of the intermediate outcomes. For example, the terminal
outcome can be college attendance, while the intermediate outcomes
are secondary school performances. In this case, we replace $Y_{T}$
with a random variable $R_{T}$ to represent the terminal outcome,
while maintaining $Y_{t}$ for $t\le T-1$ to represent the intermediate
outcomes. Analogously, $R_{T}(\boldsymbol{d})$ denotes the potential
terminal outcome. Then, the analysis in this paper can be readily
followed with the change of notation.\footnote{Extending this framework to incorporate the irreversibility of the
outcome variables discussed in Remark \ref{rem:irrev} is not straightforward.
We leave this for future research.}\end{remark}

\section{A Dynamic Structural Model and Objects of Interest\label{sec:model_and_para}}

We now introduce the main framework of this paper. Consider a\textit{
dynamic structural function} for the outcomes, where $Y_{t}$ depends
on the entire history of outcomes ($\boldsymbol{Y}^{t-1}$) as well
as the current and the entire history of treatments ($D_{t}$, $\boldsymbol{D}^{t-1}$),
and that has the form of switching regression models: For $t=1,...,T$,
\begin{align*}
Y_{t} & =\mu_{t}(\boldsymbol{Y}^{t-1},\boldsymbol{D}^{t},X_{t},U_{t}(D_{t})),
\end{align*}
where $\mu_{t}(\cdot)$ is an unknown scalar-valued function, $X_{t}$
is a set of exogenous variables, which we discuss in detail later,
and $Y_{0}$ is assumed to be exogenously determined, with $Y_{0}=0$
for convenience.\footnote{This assumption of an exogenous initial outcome is \textit{not} necessary
but only introduced to simplify our analysis; see Remark \ref{rem:initial_condi}
for alternative assumptions.} There can be other potentially endogenous covariates $W_{t}$, which
is suppressed in the model. The unobservable variable satisfies $U_{t}(D_{t})=D_{t}U_{t}(1)+(1-D_{t})U_{t}(0)$,
where $U_{t}(d_{t})$ is the ``rank variable'' that captures the
unobserved characteristics or rank, specific to treatment state $d_{t}$
(\citet{chernozhukov2005iv}). We allow $U_{it}(d_{t})$ to contain
a permanent component (i.e., individual effects) and a transitory
component.\footnote{In this case, it may make sense that the permanent component does
not depend on each $d_{t}$, but that the transitory component does.} Given this structural equation, we can express the potential outcome
$Y_{t}(\boldsymbol{d})$ using a recursive structure:
\begin{align*}
Y_{t}(\boldsymbol{d}) & =Y_{t}(\boldsymbol{d}^{t})=\mu_{t}(\boldsymbol{Y}^{t-1}(\boldsymbol{d}^{t-1}),\boldsymbol{d}^{t},X_{t},U_{t}(d_{t})),\\
 & \qquad\qquad\qquad\text{ where }\boldsymbol{Y}^{t-1}(\boldsymbol{d}^{t-1})\equiv(Y_{1}(d_{1}),Y_{2}(\boldsymbol{d}^{2}),...,Y_{t-1}(\boldsymbol{d}^{t-1})),\\
 & \vdots\\
Y_{2}(\boldsymbol{d}) & =Y_{2}(\boldsymbol{d}^{2})=\mu_{2}(Y_{1}(d_{1}),\boldsymbol{d}^{2},X_{2},U_{2}(d_{2})),\\
Y_{1}(\boldsymbol{d}) & =Y_{1}(d_{1})=\mu_{1}(Y_{0},d_{1},X_{1},U_{1}(d_{1})),
\end{align*}
where each potential outcome at time $t$ is only a function of $\boldsymbol{d}^{t}$
(not the full $\boldsymbol{d}$). This is related to the ``no-anticipation''
condition (\citet{abbring2007econometric}) or the ``consistency''
condition (\citet{robins2000marginal}), which is implied from the
structure of the model in our setting. The recursive structure provides
us with a useful interpretation of the potential outcome $Y_{t}(\boldsymbol{d})$
in a dynamic setting, and thus facilitates our identification analysis.
Imposing this structure in order to relax the sequential randomization
assumption is the trade-off we exploit. Still, we allow rich channels
in the evolution of potential outcomes, as $Y_{t}(\boldsymbol{d})$
is a function of \textit{all} the past potential outcomes whose treatment
indices are consistent with $\boldsymbol{d}$. Also, conditional on
$\boldsymbol{X}^{t}\equiv(X_{1},...,X_{t})$, the heterogeneity in
$Y_{t}(\boldsymbol{d})$ comes from the full vector $\boldsymbol{U}^{t}(\boldsymbol{d}^{t})\equiv(U_{1}(d_{1}),...,U_{t}(d_{t}))$.
By an iterative argument, we can readily show that the potential outcome
is equal to the observed outcome when the observed treatments are
consistent with the assigned regime: $Y_{t}(\boldsymbol{d})=Y_{t}$
when $\boldsymbol{D}=\boldsymbol{d}$, or equivalently, $Y_{t}=\sum_{\boldsymbol{d}\in\mathcal{D}}1\{\boldsymbol{D}=\boldsymbol{d}\}Y_{t}(\boldsymbol{d})$.

In this paper, we consider the \textit{average potential terminal
outcome}, conditional on $\boldsymbol{X}=\boldsymbol{x}$, as the
fundamental parameter of interest:
\begin{align}
E[Y_{T}(\boldsymbol{d})|\boldsymbol{X}=\boldsymbol{x}] & .\label{eq:ASF}
\end{align}
Again, we suppress that the quantity is conditional on $\boldsymbol{W}=\boldsymbol{w}$.
We also call this parameter the \textit{average recursive structural
function} (ARSF) in the terminal period, named after the recursive
structure in the model for $Y_{T}(\boldsymbol{d})$. Generally, in
defining this parameter and all others below, we can consider the
potential outcome in any time period of interest, e.g., $E[Y_{t}(\boldsymbol{d})|\boldsymbol{X}^{t}=\boldsymbol{x}^{t}]$
for any given $t$. We focus on the terminal potential outcome only
for concreteness. The knowledge of the ARSF is useful in recovering
other related parameters. 

First, we are interested in the conditional ATE:
\begin{align}
ATE(\boldsymbol{d},\tilde{\boldsymbol{d}}) & \equiv E[Y_{T}(\boldsymbol{d})-Y_{T}(\tilde{\boldsymbol{d}})|\boldsymbol{X}=\boldsymbol{x}]\label{eq:ATE}
\end{align}
for two different regimes, $\boldsymbol{d}$ and $\tilde{\boldsymbol{d}}$.
For example, one may be interested in comparing more versus less consistent
treatment sequences, or earlier versus later treatments. 

Second, we consider the \textit{optimal treatment regime}:
\begin{align}
\boldsymbol{d}^{*}(w_{0}) & =\arg\max_{\boldsymbol{d}\in\mathcal{D}}E[Y_{T}(\boldsymbol{d})|W_{0}=w_{0}]\label{eq:d*}
\end{align}
with $\left|\mathcal{D}\right|\le2^{T}$, where $W_{0}$ is a vector
of pre-treatment covariates in $W_{t}=(W_{0},W_{1t})$. That is, we
are interested in a treatment regime that delivers the maximum expected
potential outcome, conditional on characteristics $W_{0}=w_{0}$.
Notice that, in a static model, the identification of $\boldsymbol{d}^{*}$
is equivalent to the identification of the sign of the static ATE,
which is the information typically sought from a policy point of view.
One can view $\boldsymbol{d}^{*}$ as a natural extension of this
information to a dynamic setting, which is identified by establishing
the signs of \textit{all} possible ATE's defined as in \eqref{eq:ATE},
or equivalently, by ordering all the possible ARSF's. The optimal
regime may serve as a guideline in developing future policies. Moreover,
it may be a realistic goal for a social planner to identify this kind
of scheme that maximizes the average benefit, because it may be too
costly to find a customized treatment scheme for every individual.
Yet, the optimal regime is customized up to observed pre-treatment
characteristics, as it is a function of $w_{0}$. Given $\boldsymbol{d}^{*}(w_{0})$,
we may be interested in $E[Y_{T}(\boldsymbol{d}^{*}(w_{0}))]$ or
the ATE for the effect of $\boldsymbol{d}^{*}(w_{0})$ relative to
another treatment sequence (e.g., the second best). More ambitious
than the identification of $\boldsymbol{d}^{*}(w_{0})$ may be recovering
an optimal regime based on a cost\textendash benefit analysis, granting
than each $d_{t}$ can be costly:
\begin{align}
\boldsymbol{d}^{\dagger}(w_{0}) & =\arg\max_{\boldsymbol{d}\in\mathcal{D}}\Pi(\boldsymbol{d};w_{0}),\label{eq:d^dag}
\end{align}
where
\begin{align*}
\Pi(\boldsymbol{d};w_{0}) & \equiv wE[Y_{T}(\boldsymbol{d})|W_{0}=w_{0}]-\tilde{w}\sum_{t=1}^{T}d_{t}\quad\text{or}\quad\Pi(\boldsymbol{d};w_{0})\equiv\sum_{t=1}^{T}w_{t}E[Y_{t}(\boldsymbol{d})|W_{0}=w_{0}]-\sum_{t=1}^{T}\tilde{w}_{t}d_{t}
\end{align*}
with $(w,\tilde{w})$ and $(\boldsymbol{w},\tilde{\boldsymbol{w}})$
being predetermined weights. The latter objective function concerns
the weighted sum of the average potential outcomes throughout the
entire period, less the cost of treatments. Note that establishing
the signs of ATE's will not identify $\boldsymbol{d}^{\dagger}$,
and a stronger identification result becomes important, i.e., the
point identification of $E[Y_{T}(\boldsymbol{d})|W_{0}=w_{0}]$ for
all $\boldsymbol{d}$ (or $E[Y_{t}(\boldsymbol{d})|W_{0}=w_{0}]$
for all $t$ and $\boldsymbol{d}$). 

Lastly, we are interested in the \textit{transition-specific ATE}:
\begin{align}
E[Y_{T}(\boldsymbol{d})|Y_{T-1}(\boldsymbol{d})=y_{T-1},\boldsymbol{X}=\boldsymbol{x}]-E[Y_{T}(\tilde{\boldsymbol{d}})|Y_{T-1}(\tilde{\boldsymbol{d}})=y_{T-1},\boldsymbol{X}=\boldsymbol{x}]\label{eq:ATET}
\end{align}
for two different $\boldsymbol{d}$ and $\tilde{\boldsymbol{d}}$.
The knowledge of the ARSF does not directly recover this parameter,
but the identification of it (and its more general form introduced
later) can be paralleled by the analysis for the ARSF and ATE. 

In order to facilitate identification of the parameters of interest
without assuming sequential randomization, we introduce a sequence
of selection equations for the binary endogenous treatments, where
$D_{t}$ depends on the entire history of outcomes and treatments
($\boldsymbol{Y}^{t-1}$ and $\boldsymbol{D}^{t-1}$): For $t=1,...,T$,
\begin{align*}
D_{t} & =1\{\pi_{t}(\boldsymbol{Y}^{t-1},\boldsymbol{D}^{t-1},Z_{t})\ge V_{t}\},
\end{align*}
where $\pi_{t}(\cdot)$ is an unknown scalar-valued function, $Z_{t}$
is the period-specific instruments, $V_{t}$ is the unobservable variable
that may contain permanent and transitory components, and $D_{0}$
is assumed to be exogenously given as $D_{0}=0$.\footnote{This is an alternative to simply assuming there is no treatment at
$t=0$. We maintain the current assumption to avoid additional definitions
for $\pi_{1}(\cdot)$ and other relevant objects.} This dynamic selection process represents the agent's endogenous
choices over time, e.g., as a result of learning or other optimal
behaviors. However, the nonparametric threshold-crossing structure
posits a minimal notion of optimality for the agent. We take an agnostic
approach by avoiding strong assumptions of the standard dynamic economic
models pioneered by \citet{rust1987optimal}, such as forward looking
behaviors and being able to compute a present value discounted flow
of utilities. If we are to maintain the assumption of rational agents,
the selection model can be viewed as a reduced-form approximation
of a solution to a dynamic programming problem. Lastly, due to the
dynamic structure, this selection equation does not necessarily imply
the monotonicity assumption of \citet{imbens1994identification} or
vice versa.

To simplify the exposition, we consider binary $Y_{t}$ and impose
weak separability in the outcome equation as in the treatment equation.
The binary outcome is \textit{not} necessary for the result of this
paper, and the analysis can be easily extended to the case of continuous
or censored $Y_{t}$, maintaining weak separability; see Remark \ref{rem:contiY}.
Then, the full model can be summarized as
\begin{align}
Y_{t} & =1\{\mu_{t}(\boldsymbol{Y}^{t-1},\boldsymbol{D}^{t},X_{t})\ge U_{t}(D_{t})\},\label{eq:model1}\\
D_{t} & =1\{\pi_{t}(\boldsymbol{Y}^{t-1},\boldsymbol{D}^{t-1},Z_{t})\ge V_{t}\}.\label{eq:model2}
\end{align}
In this model, the observable variables are $(\boldsymbol{Y},\boldsymbol{D},\boldsymbol{X},\boldsymbol{Z})$.
All other covariates $W_{t}$ are suppressed in the equations for
simplicity of exposition. Importantly, in this model, the joint distribution
of the unobservable variables $(\boldsymbol{U}(\boldsymbol{d}),\boldsymbol{V})$
for given $\boldsymbol{d}$ is not specified, in that $U_{t}(d_{t})$
and $V_{t'}$ for any $t,t'$ are allowed to be arbitrarily correlated
to each other (allowing endogeneity) as well as within themselves
across time (allowing serial correlation, e.g., via time-invariant
individual effects). Note that, because we allow an arbitrary form
of persistence in the unobservables and the dependence of $Y_{t}$
and $D_{t}$ on the entire history, $(Y_{t},D_{t})$ is \textit{not}
a Markov process even after conditioning on the observables. This
is in contrast to the standard dynamic economic models, where conditional
independence assumptions or Markovian unobservables are commonly introduced.
By considering the nonparametric index functions that depend on $t$,
we also avoid other strong assumptions on parametric functional forms
or time homogeneity.

\begin{remark}[\textbf{Irreversibility}---continued]\label{rem:irrev2}A
process that satisfies $D_{t}=1$ if $D_{t-1}=1$ is consistent with
having a structural function that satisfies $\pi_{t}(\boldsymbol{y}^{t-1},\boldsymbol{d}^{t-1},z_{t})=+\infty$
if $d_{t-1}=1$. Similarly, processes that satisfy $Y_{t}=1$ and
$D_{t}=0$ if $Y_{t-1}=1$ are consistent with $\mu_{t}(\boldsymbol{y}^{t-1},\boldsymbol{d}^{t},x_{t})=+\infty$
and $\pi_{t}(\boldsymbol{y}^{t-1},\boldsymbol{d}^{t-1},z_{t})=-\infty$
if $y_{t-1}=1$. This implies that $Y_{t}(\boldsymbol{d}^{t})=1$
for any $d_{t}$ if $Y_{t-1}(\boldsymbol{d}^{t-1})=1$. When $Y_{t}$
is irreversible, the ARSF $E[Y_{T}(\boldsymbol{d})|X]$ can be interpreted
as (one minus) a potential survival rate. An important caveat is that,
with irreversible $Y_{t}$, the ATE we define contains not only the
treatment effect (the intensive margin) but also the effect on dynamic
selection (the extensive margin), and the parameter may or may not
be of interest depending on the application.\end{remark}

\begin{remark}[\textbf{Terminal outcome of a different kind}---continued]\label{rem:term2}When
we replace $Y_{T}$ with $R_{T}$ to represent a terminal outcome
of a different kind, we assume that the model \eqref{eq:model1} is
only satisfied for $t\le T-1$ and introduce $R_{T}=1\{\mu_{T}(\boldsymbol{Y}^{T-1},\boldsymbol{D}^{T},X_{T})\ge U_{T}(D_{T})\}$
as the terminal structural function. The potential terminal outcome
$R_{T}(\boldsymbol{d})$ can accordingly be expressed using the structural
functions for $(Y_{1},...,Y_{T-1},R_{T})$. The ARSF is written as
$E[R_{T}(\boldsymbol{d})|X]$, and the other parameters can be defined
accordingly.\end{remark}

\begin{remark}[\textbf{Non-binary $Y_t$}]\label{rem:contiY}Even
though we focus on binary $Y_{t}$ in this paper, we can obtain similar
identification results with continuous $Y_{t}$ or limited dependent
variable $Y_{t}$, by maintaining a general weak separability structure:
$Y_{t}=m_{t}(\mu_{t}(\boldsymbol{Y}^{t-1},\boldsymbol{D}^{t},X_{t}),U_{t}(D_{t}))$.
As in the static settings of \citet{VY07} and \citet{Han2018}, we
impose an assumption that guarantees certain monotonicity of each
period's average structural function with respect to the index $\mu_{t}$:
For each $t$, $E[m_{t}(\mu_{t},U_{t}(d_{t}))|\boldsymbol{V}^{t},\boldsymbol{U}^{t-1}]$
is strictly monotonic in $\mu_{t}$. Examples of the nonparametric
model $m_{t}(\mu_{t}(\boldsymbol{y}^{t-1},\boldsymbol{d}^{t},x_{t}),u_{t})$
that satisfies this assumption are additively separable models or
their transformation models, censored regression models, and threshold
crossing models as in \eqref{eq:model1}; see \citet{VY07} for more
discussions.\end{remark}

\section{Motivating Example\label{sec:Motivating-Example}}

A multi-period experiment with imperfect compliance is one motivating
example of this paper's setup. Multi-period experiments are common
in clinical trials, such as in the Fast Track Prevention Program (\citet*{conduct1992developmental}),
the Elderly Program randomized trial for the Systolic Hypertension
(\citet{the1988rationale}), and the AIDS Clinical Trial Group\footnote{The AIDS Clinical Trials Group (https://actgnetwork.org) is one of
the largest HIV clinical trials organizations in the world.}; also see the biostatistics literature referenced in the introduction
for other examples. For instance, the Fast Track Prevention Program
is a randomized trial to prevent conduct disorders and drug use in
children at risk. Interventions are taken place at the end of each
semester starting from first grade, by means of home visits and teacher
consultations. In household visits, for example, it is reported that
assignment deviation occurs for nearly 50\% of the intervention children.
\citet{murphy2001marginal} focus on the effect of treatment had there
been no deviation, i.e., the intention-to-treat parameters. In this
paper, we recover the average treatment effect parameters allowing
for this type of imperfect compliance.

Based on to these clinical trials, we consider the following stylized
example for the structural model of this paper. A clinical research
organization is interested in improving patients' symptoms ($Y_{t}$),
and runs an experiment of randomly assigning treatments at each $t$
($Z_{t}$). Based on the assignment, each patient decides whether
or not to receive the treatment ($D_{t}$) by being a complier, defier,
always-taker or never-taker. This information can be collected via
a fidelity assessment as in the Fast Track Prevention Program. In
making the compliance decision, the patient has a habit ($\boldsymbol{D}^{t-1}$)
and takes into account her past symptoms ($\boldsymbol{Y}^{t-1}$).
The current symptom ($Y_{t}$) is formed based on the past symptoms
($\boldsymbol{Y}^{t-1}$), the current and past treatment take-ups
($\boldsymbol{D}^{t}$), and other symptom-influencing factors ($X_{t}$)
occurring at time $t$. As described in detail in the next section,
we assume that patients cannot fully predict $X_{t}$ when making
treatment decisions $D_{t}$. For patients with potential respiratory
diseases, temporal variation in air quality can be such a variable.
In the Fast Track Prevention Program, the average performance measure
of non-risk peers randomly assigned every academic year can be a candidate.

\section{Main Identification Analysis\label{sec:ID}}

We first identify the ARSF's, i.e., $E[Y_{t}(\boldsymbol{d})|\boldsymbol{X}^{t}]$
for every $\boldsymbol{d}$ and $t$, which will then be used to identify
the ATE's and the optimal regimes $\boldsymbol{d}^{*}$ and $\boldsymbol{d}^{\dagger}$.
We maintain the following assumptions on $(\boldsymbol{Z},\boldsymbol{X})$
and $(\boldsymbol{U}(\boldsymbol{d}),\boldsymbol{V})$ for every $\boldsymbol{d}$.
These assumptions are written for the identification of $E[Y_{T}(\boldsymbol{d})|\boldsymbol{X}]$,
and are sufficient but not necessary for the identification of $E[Y_{t}(\boldsymbol{d})|\boldsymbol{X}^{t}]$
for $t\le T-1$.\begin{asC}The distribution of $(\boldsymbol{U}(\boldsymbol{d}),\boldsymbol{V})$
has strictly positive density with respect to Lebesgue measure on
$\mathbb{R}^{2T}$.\end{asC}\begin{asSX}$(\boldsymbol{Z},\boldsymbol{X})$
and $(\boldsymbol{U}(\boldsymbol{d}),\boldsymbol{V})$ are independent.\end{asSX}Assumption
C is a regularity condition to ensure the smoothness of relevant conditional
probabilities. Assumption SX imposes strict exogeneity, which is a
simple sufficient condition for necessary requirements we need for
identification; see Remark \ref{rem:about_SX}. It is implicit that
the independence is conditional on the covariates suppressed in the
model. Just as the treatments $\boldsymbol{D}$, these covariates
may be correlated with the individual effects contained in $(\boldsymbol{U}(\boldsymbol{d}),\boldsymbol{V})$.
The variable $Z_{t}$ denotes the standard excluded instruments, which
is allowed to be binary. A leading example is a sequence of randomized
treatment assignments. Other examples include sequential policy shocks.
In addition to $Z_{t}$, we introduce exogenous variables $X_{t}$
in the outcome equation \eqref{eq:model1}, that are excluded from
the selection equation \eqref{eq:model2}. We make a behavioral/information
assumption that there are outcome-determining factors that the agent
cannot fully anticipate when making a treatment decision. Continuing
with the stylized example in Section \ref{sec:Motivating-Example},
when $D_{t}$ is a compliance choice that a patient makes at the $t$-th
visit to the clinical facility, $Y_{t-1}$ may be the symptom measured
prior to the decision during the \textit{same} visit. Then $Y_{t}$
is the symptom measured upon the next visit, which may create enough
time gap to prevent the patient from predicting $X_{t}$.\footnote{In a static scenario, \citet{Han2018} motivate this reverse exclusion
restriction using the notion of externalities. In their setting where
multiple treatments are strategically chosen (e.g., firms' entry decisions),
factors that determine the outcome (e.g., pollution) are assumed not
to appear in the firms' payoff functions. } Note that $(Z_{t},X_{t})$ are assumed to be excluded from the
outcome and treatment equations of all other periods as well. Next,
we introduce a sequential version of the rank similarity assumption
(\citet{chernozhukov2005iv}):\begin{asRS}For each $t$ and $\boldsymbol{d}_{-t}$,
$\boldsymbol{U}(1,\boldsymbol{d}_{-t})$ and $\boldsymbol{U}(0,\boldsymbol{d}_{-t})$
are identically distributed, conditional on $\boldsymbol{V}^{t}$
and $(\boldsymbol{Z},\boldsymbol{X})$.\end{asRS}Rank invariance
(i.e., $\{\boldsymbol{U}(\boldsymbol{d})\}_{\boldsymbol{d}}$ being
equal to each other) is particularly restrictive in the multi-period
context, because it requires that the same rank be realized across
$2^{T}$ different treatment states. Significantly weaker than the
rank invariance would be a joint rank similarity assumption that $\boldsymbol{U}(\boldsymbol{d})$'s
are identically distributed across $2^{T}$ states (conditional on
the observables and treatment unobservables). This allows an individual
to have different realized ranks across different $\boldsymbol{d}$'s.
Assumption RS, which we call \textit{sequential rank similarity},
relaxes this even further by only requiring that $\boldsymbol{U}(1,\boldsymbol{d}_{-t})$
and $\boldsymbol{U}(0,\boldsymbol{d}_{-t})$ are identically distributed
instead. That is, the assumption requires that, within individuals
with the same observed characteristics and history of the treatment
unobservables, the joint distributions of the ranks are identical
between just two states that differ by $d_{t}=1$ and $0$.\footnote{In fact, we can further relax Assumption RS by allowing $U_{t}(d_{t})$
to be a function of $x_{t}$ from the outset; see Remark \ref{rem:rank_with_x}.}

Now, we are ready to derive a period-specific result. Define the following
period-specific quantity directly identified from the data, i.e.,
from the distribution of $(\boldsymbol{Y},\boldsymbol{D},\boldsymbol{X},\boldsymbol{Z})$:
\begin{align*}
 & h_{t}(z_{t},\tilde{z}_{t},x_{t},\tilde{x}_{t};\boldsymbol{z}^{t-1},\boldsymbol{x}^{t-1},\boldsymbol{d}^{t-1},\boldsymbol{y}^{t-1})\\
\equiv & \Pr[Y_{t}=1,D_{t}=1|\boldsymbol{z}^{t},\boldsymbol{x}^{t},\boldsymbol{d}^{t-1},\boldsymbol{y}^{t-1}]+\Pr[Y_{t}=1,D_{t}=0|\boldsymbol{z}^{t},\tilde{x}_{t},\boldsymbol{x}^{t-1},\boldsymbol{d}^{t-1},\boldsymbol{y}^{t-1}]\\
 & -\Pr[Y_{t}=1,D_{t}=1|\tilde{z}_{t},\boldsymbol{z}^{t-1},\boldsymbol{x}^{t},\boldsymbol{d}^{t-1},\boldsymbol{y}^{t-1}]-\Pr[Y_{t}=1,D_{t}=0|\tilde{z}_{t},\boldsymbol{z}^{t-1},\tilde{x}_{t},\boldsymbol{x}^{t-1},\boldsymbol{d}^{t-1},\boldsymbol{y}^{t-1}]
\end{align*}
for $t\ge1$, where $(\boldsymbol{Z}^{0},\boldsymbol{X}^{0},\boldsymbol{D}^{0},Y_{0})$
is understood to mean that there is no conditioning.\begin{lemma}\label{lem:sign_of_period_ATE}Suppose
Assumptions C, SX and RS hold. For each $t$ and $(\boldsymbol{z}^{t-1},\boldsymbol{x}^{t-1},\boldsymbol{d}^{t-1},\boldsymbol{y}^{t-1})$,
suppose $z_{t}$ and $\tilde{z}_{t}$ are such that
\begin{align}
\Pr[D_{t}=1|\boldsymbol{z}^{t},\boldsymbol{x}^{t-1},\boldsymbol{d}^{t-1},\boldsymbol{y}^{t-1}] & \neq\Pr[D_{t}=1|\tilde{z}_{t},\boldsymbol{z}^{t-1},\boldsymbol{x}^{t-1},\boldsymbol{d}^{t-1},\boldsymbol{y}^{t-1}].\label{eq:rf_condi}
\end{align}
Then, for given $(x_{t},\tilde{x}_{t})$, the sign of $h_{t}(z_{t},\tilde{z}_{t},x_{t},\tilde{x}_{t};\boldsymbol{z}^{t-1},\boldsymbol{x}^{t-1},\boldsymbol{d}^{t-1},\boldsymbol{y}^{t-1})$
is equal to the sign of $\mu_{t}(\boldsymbol{y}^{t-1},\boldsymbol{d}^{t-1},1,x_{t})-\mu_{t}(\boldsymbol{y}^{t-1},\boldsymbol{d}^{t-1},0,\tilde{x}_{t})$.\end{lemma}Without
relying on further assumptions, the sign of $\mu_{t}(\boldsymbol{y}^{t-1},\boldsymbol{d}^{t-1},1,x_{t})-\mu_{t}(\boldsymbol{y}^{t-1},\boldsymbol{d}^{t-1},0,\tilde{x}_{t})$
itself is already useful for calculating bounds on the ARSF's and
thus on the ATE's; we discuss the partial identification in Section
\ref{sec:partial_ID}.

For the analysis of this paper which deals with a dynamic model, it
is convenient to define the $\boldsymbol{U}$-set and $\boldsymbol{V}$-set,
namely the sets of histories of the unobservable variables that determine
the outcomes and treatments, respectively. To focus our attention
on the dependence of the potential outcomes on the unobservables,
we iteratively define the potential outcome given $(\boldsymbol{d},\boldsymbol{x})$
as
\begin{align*}
Y_{t}(\boldsymbol{d}^{t},\boldsymbol{x}^{t}) & \equiv1\{\mu_{t}(\boldsymbol{Y}^{t-1}(\boldsymbol{d}^{t-1},\boldsymbol{x}^{t-1}),\boldsymbol{d}^{t},x_{t})\ge U_{t}(d_{t})\}
\end{align*}
for $t\ge2$, with $Y_{1}(d_{1},x_{1})=1\{\mu_{1}(0,d_{1},x_{1})\ge U_{1}(d_{1})\}$.
Now, define the set of $\boldsymbol{U}^{t}(\boldsymbol{d}^{t})$ as
\begin{align*}
\mathcal{U}^{t}(\boldsymbol{d}^{t},\boldsymbol{y}^{t})\equiv\mathcal{U}^{t}(\boldsymbol{d}^{t},\boldsymbol{y}^{t};\boldsymbol{x}^{t}) & \equiv\{\boldsymbol{U}^{t}(\boldsymbol{d}^{t}):y_{s}=Y_{s}(\boldsymbol{d}^{s},\boldsymbol{x}^{s})\text{ for all }s\le t\}.
\end{align*}
for $t\ge1$. Then, $\boldsymbol{Y}^{t}=\boldsymbol{y}^{t}$ if and
only if $\boldsymbol{U}^{t}(\boldsymbol{d}^{t})\in\mathcal{U}^{t}(\boldsymbol{d}^{t},\boldsymbol{y}^{t};\boldsymbol{x}^{t})$,
conditional on $(\boldsymbol{D}^{t},\boldsymbol{X}^{t})=(\boldsymbol{d}^{t},\boldsymbol{x}^{t})$.
The $\boldsymbol{V}$-set $\mathcal{V}^{t}(\boldsymbol{d}^{t},\boldsymbol{u}^{t-1})\equiv\mathcal{V}^{t}(\boldsymbol{d}^{t},\boldsymbol{u}^{t-1};\boldsymbol{z}^{t},\boldsymbol{x}^{t-1})$
is similarly defined within the proof of Lemma \ref{lem:sign_of_period_ATE}
in the Appendix. Then, $\boldsymbol{D}^{t}=\boldsymbol{d}^{t}$
if and only if $\boldsymbol{V}^{t}\in\mathcal{V}^{t}(\boldsymbol{d}^{t},\boldsymbol{U}^{t-1}(\boldsymbol{d}^{t-1}))$,
conditional on $(\boldsymbol{Z}^{t},\boldsymbol{X}^{t-1})=(\boldsymbol{z}^{t},\boldsymbol{x}^{t-1})$.
Given these sets, what we show in the proof of this lemma is that,
under Assumptions C and SX,
\begin{align*}
 & h_{t}(z_{t},\tilde{z}_{t},x_{t},\tilde{x}_{t};\boldsymbol{z}^{t-1},\boldsymbol{x}^{t-1},\boldsymbol{d}^{t-1},\boldsymbol{y}^{t-1})\\
= & \Pr[U_{t}(1)\le\mu_{t}(\boldsymbol{y}^{t-1},\boldsymbol{d}^{t-1},1,x_{t}),\tilde{\pi}_{t}\le V_{t}\le\pi_{t}|\mathcal{V}^{t-1}(\boldsymbol{d}^{t-1},\boldsymbol{U}^{t-2}(\boldsymbol{d}^{t-2})),\mathcal{U}^{t-1}(\boldsymbol{d}^{t-1},\boldsymbol{y}^{t-1})]\\
 & -\Pr[U_{t}(0)\le\mu_{t}(\boldsymbol{y}^{t-1},\boldsymbol{d}^{t-1},0,\tilde{x}_{t}),\tilde{\pi}_{t}\le V_{t}\le\pi_{t}|\mathcal{V}^{t-1}(\boldsymbol{d}^{t-1},\boldsymbol{U}^{t-2}(\boldsymbol{d}^{t-2})),\mathcal{U}^{t-1}(\boldsymbol{d}^{t-1},\boldsymbol{y}^{t-1})],
\end{align*}
the sign of which identifies the sign of $\mu_{t}(\boldsymbol{y}^{t-1},\boldsymbol{d}^{t-1},1,x_{t})-\mu_{t}(\boldsymbol{y}^{t-1},\boldsymbol{d}^{t-1},0,\tilde{x}_{t})$
by Assumption RS. For example, when this quantity is zero, then $\mu_{t}(\boldsymbol{y}^{t-1},\boldsymbol{d}^{t-1},1,x_{t})-\mu_{t}(\boldsymbol{y}^{t-1},\boldsymbol{d}^{t-1},0,\tilde{x}_{t})=0$.

For the point identification of the ARSF's, the final assumption we
introduce concerns the variation of the exogenous variables $(\boldsymbol{Z},\boldsymbol{X})$.
Define the following sets:
\begin{align}
\mathcal{S}_{t}(\boldsymbol{d}^{t},\boldsymbol{y}^{t-1}) & \equiv\left\{ (x_{t},\tilde{x}_{t}):\mu_{t}(\boldsymbol{y}^{t-1},\boldsymbol{d}^{t},x_{t})=\mu_{t}(\boldsymbol{y}^{t-1},\boldsymbol{d}^{t-1},\tilde{d}_{t},\tilde{x}_{t})\text{ for }\tilde{d}_{t}\neq d_{t}\right\} ,\label{eq:set1}\\
\mathcal{T}_{t}(\boldsymbol{x}_{-t},\boldsymbol{z}_{-t}) & \equiv\left\{ (x_{t},\tilde{x}_{t}):\exists(z_{t},\tilde{z}_{t})\text{ such that \eqref{eq:rf_condi} holds and}\right.\nonumber \\
 & \qquad\left.(x_{t},z_{t}),(\tilde{x}_{t},z_{t}),(x_{t},\tilde{z}_{t}),(\tilde{x}_{t},\tilde{z}_{t})\in\text{Supp}(X_{t},Z_{t}|\boldsymbol{x}_{-t},\boldsymbol{z}_{-t})\right\} ,\label{eq:set2}\\
\mathcal{X}_{t}(\boldsymbol{d}^{t},\boldsymbol{y}^{t-1};\boldsymbol{x}_{-t},\boldsymbol{z}_{-t}) & \equiv\left\{ x_{t}:\exists\tilde{x}_{t}\text{ with }(x_{t},\tilde{x}_{t})\in\mathcal{S}_{t}(\boldsymbol{d}^{t},\boldsymbol{y}^{t-1})\cap\mathcal{T}_{t}(\boldsymbol{x}_{-t},\boldsymbol{z}_{-t})\right\} ,\label{eq:set3}\\
\mathcal{X}_{t}(\boldsymbol{d}^{t};\boldsymbol{x}_{-t},\boldsymbol{z}_{-t}) & \equiv\bigcap_{\boldsymbol{y}^{t-1}}\mathcal{X}_{t}(\boldsymbol{d}^{t},\boldsymbol{y}^{t-1};\boldsymbol{x}_{-t},\boldsymbol{z}_{-t}),\label{eq:set4}
\end{align}
where \eqref{eq:set1} is related to the sufficient variation of $X_{t}$
and \eqref{eq:set2} is related to the rectangular variation of $(X_{t},Z_{t})$.
\begin{asSP}For each $t$ and $\boldsymbol{d}^{t}$, $\Pr[X_{t}\in\mathcal{X}_{t}(\boldsymbol{d}^{t};\boldsymbol{x}_{-t},\boldsymbol{z}_{-t})|\boldsymbol{x}_{-t},\boldsymbol{z}_{-t}]>0$
almost everywhere.\end{asSP}This assumption requires that $X_{t}$
varies sufficiently to achieve $\mu_{t}(\boldsymbol{y}^{t-1},\boldsymbol{d}^{t},x_{t})=\mu_{t}(\boldsymbol{y}^{t-1},\boldsymbol{d}^{t-1},\tilde{d}_{t},\tilde{x}_{t})$,
while holding $Z_{t}$ to be $z_{t}$ and $\tilde{z}_{t}$, respectively,
conditional on $(\boldsymbol{X}_{-t},\boldsymbol{Z}_{-t})$. This
is a dynamic version of the support assumption found in \citet{VY07}.\footnote{In our setting, it is possible that $\mathcal{X}_{t}(\boldsymbol{d}^{t};\boldsymbol{x}_{-t},\boldsymbol{z}_{-t})$
is nonempty even when $Z_{t}$ is discrete, as long as $X_{t}$ contains
continuous elements with sufficient support (\citet{VY07}). In all
these works, including the present one, the support requirement is
conditional on the exogenous variables in other periods; see also
\citet{cameron1998life}.} Although Assumption SP requires sufficient rectangular variation
in $(X_{t},Z_{t})$, it clearly differs from the large variation
assumptions in, e.g., \citet{heckman2007dynamic} and \citet{heckman2016dynamic}.
These papers employ identification-at-infinity arguments in each period
that the support of explained variation (i.e., $\mu_{t}(\cdot)$ in
our notation) is no smaller than the support of unobservables. On
the other hand, Assumption SP only requires the existence of variation
that equates $\mu_{t}(\cdot)$ for two different values of $D_{t}$.
Apparently, this is trivially satisfied with the former assumption
of large support. Note that even though Assumption SP seems to be
written in terms of the unknown object $\mu_{t}(\cdot)$, it is testable
because the sets defined above have empirical analogs, according to
Lemma \ref{lem:sign_of_period_ATE}. Let $\mathcal{X}_{t}(\boldsymbol{d}^{t};\boldsymbol{x}_{-t})\equiv\{x_{t}:x_{t}\in\mathcal{X}_{t}(\boldsymbol{d}^{t};\boldsymbol{x}_{-t},\boldsymbol{z}_{-t})\text{ for some }\boldsymbol{z}_{-t}\in\text{Supp}(\boldsymbol{Z}_{-t}|\boldsymbol{x}_{-t})\}$
and $\mathcal{X}(\boldsymbol{d})\equiv\{\boldsymbol{x}:x_{t}\in\mathcal{X}_{t}(\boldsymbol{d}^{t};\boldsymbol{x}_{-t})\text{ for some }(x_{t+1},...,x_{T}),\text{ for }t\ge1\}$,
which sequentially collect $x_{t}\in\mathcal{X}_{t}(\boldsymbol{d}^{t};\boldsymbol{x}_{-t},\boldsymbol{z}_{-t})$
for all $t$. We are now ready to state the main identification result.\begin{theorem}\label{thm:ASF}Under
Assumptions C, SX, RS and SP, $E[Y_{T}(\boldsymbol{d})|\boldsymbol{x}]$
is identified for $\boldsymbol{d}\in\mathcal{D}$ and $\boldsymbol{x}\in\mathcal{X}(\boldsymbol{d})$.\end{theorem}Based
on Theorem \ref{thm:ASF}, we can identify the ATE's. Since the identification
of all $E[Y_{t}(\boldsymbol{d})|\boldsymbol{x}^{t}]$'s can be shown
analogously to Theorem \ref{thm:ASF}, we can identify the optimal
treatment regimes $\boldsymbol{d}^{*}(\boldsymbol{x})$ and $\boldsymbol{d}^{\dagger}(\boldsymbol{x})$
as well.\begin{corollary}\label{cor:ID}Under Assumptions C, SX,
RS and SP, $ATE(\boldsymbol{d},\tilde{\boldsymbol{d}})$ is identified
for $\boldsymbol{d},\tilde{\boldsymbol{d}}\in\mathcal{D}$ and $\boldsymbol{x}\in\mathcal{X}(\boldsymbol{d})\cap\mathcal{X}(\tilde{\boldsymbol{d}})$,
and $\boldsymbol{d}^{*}(w_{0})$ and $\boldsymbol{d}^{\dagger}(w_{0})$
are identified for $w_{0}$ in its support $\mathcal{W}_{0}$.\end{corollary}

We sketch the identification analysis here; the full proof of Theorem
\ref{thm:ASF} is found in the Appendix. We consider the identification
of $E[Y_{T}(\boldsymbol{d})|\boldsymbol{x},\boldsymbol{z}]$, since
$E[Y_{T}(\boldsymbol{d})|\boldsymbol{x}]=E[Y_{T}(\boldsymbol{d})|\boldsymbol{x},\boldsymbol{z}]$
by Assumption SX.\footnote{When we are to identify the average potential outcome at $t$ instead,
the conditioning variables we use are the vectors of exogenous variables
up to $t$, i.e., $E[Y_{t}(\boldsymbol{d}^{t})|\boldsymbol{x}^{t},\boldsymbol{z}^{t}]$.
Then the entire proof can be easily modified based on this expression.} As the first step of identifying $E[Y_{T}(\boldsymbol{d})|\boldsymbol{x},\boldsymbol{z}]$
for given $\boldsymbol{d}=(d_{1},...,d_{T})$, $\boldsymbol{x}=(x_{1},...,x_{T})$
and $\boldsymbol{z}=(z_{1},...,z_{T})$, we apply the result of Lemma
\ref{lem:sign_of_period_ATE}. Fix $t\ge2$ and $\boldsymbol{y}^{t-1}\in\{0,1\}^{t-1}$.
Suppose $x_{t}'$ is such that $\mu_{t}(\boldsymbol{y}^{t-1},\boldsymbol{d}^{t},x_{t})=\mu_{t}(\boldsymbol{y}^{t-1},\boldsymbol{d}^{t-1},d_{t}',x_{t}')$
with $d_{t}'\neq d_{t}$ by applying Lemma \ref{lem:sign_of_period_ATE}.
The existence of $x_{t}'$ is guaranteed by Assumption SP, as $x_{t}\in\mathcal{X}_{t}(\boldsymbol{d}^{t},\boldsymbol{y}^{t-1};\boldsymbol{x}_{-t},\boldsymbol{z}_{-t})\subset\mathcal{X}_{t}(\boldsymbol{d}^{t};\boldsymbol{x}_{-t},\boldsymbol{z}_{-t})$.
The implication of $\mu_{t}(\boldsymbol{y}^{t-1},\boldsymbol{d}^{t},x_{t})=\mu_{t}(\boldsymbol{y}^{t-1},\boldsymbol{d}^{t-1},d_{t}',x_{t}')$
for relevant $\boldsymbol{U}$-sets is as follows: Analogous to the
$\boldsymbol{U}$-set defined earlier, define
\begin{align*}
\mathcal{U}^{t}(\boldsymbol{d}^{t},y_{t})\equiv\mathcal{U}^{t}(\boldsymbol{d}^{t},y_{t};\boldsymbol{x}^{t}) & \equiv\{\boldsymbol{U}^{t}(\boldsymbol{d}^{t}):y_{t}=Y_{t}(\boldsymbol{d}^{t},\boldsymbol{x}^{t})\}.
\end{align*}
Then, by definition, $\boldsymbol{U}\in\mathcal{U}(\boldsymbol{d},y_{T};\boldsymbol{x})$
is equivalent to $\boldsymbol{U}\in\mathcal{U}^{T}(d_{t}',\boldsymbol{d}_{-t},y_{T};x_{t}',\boldsymbol{x}_{-t})$
\textit{conditional on} $\boldsymbol{Y}^{t-1}(\boldsymbol{d}^{t-1},\boldsymbol{x}^{t-1})=\boldsymbol{y}^{t-1}$
for all $\boldsymbol{x}_{-t}$ and $\boldsymbol{d}_{-t}$.\footnote{The subsequent analysis is substantially simplified when $\mu_{t}(y_{t-1},d_{t},x_{t})=\mu_{t}(y_{t-1},d_{t}',x_{t}')$
is satisfied \textit{for all} $y_{t-1}$, but this situation is unlikely
to occur. Therefore, it is important to condition on $Y_{t-1}(\boldsymbol{d}^{t-1},\boldsymbol{x}^{t-1})=y_{t-1}$
in the analysis.} Based on this result, we equate the unobserved quantity $E[Y_{T}(\boldsymbol{d})|\boldsymbol{x},\boldsymbol{z},\boldsymbol{y}^{t-1},\boldsymbol{d}^{t-1},d_{t}']$
with a quantity that partly matches the assigned treatment and the
observed treatment as follows. First, we can show that
\begin{align*}
 & E[Y_{T}(\boldsymbol{d})|\boldsymbol{x},\boldsymbol{z},\boldsymbol{y}^{t-1},\boldsymbol{d}^{t-1},d_{t}']\\
= & \Pr\left[\left.\begin{array}{c}
\\
\boldsymbol{U}(\boldsymbol{d})\in\mathcal{U}^{T}(\boldsymbol{d},1;\boldsymbol{x})\\
\\
\end{array}\right|\begin{array}{c}
\boldsymbol{U}^{t-1}(\boldsymbol{d}^{t-1})\in\mathcal{U}^{t-1}(\boldsymbol{d}^{t-1},\boldsymbol{y}^{t-1}),\\
\boldsymbol{V}^{t}\in\mathcal{V}^{t}(\boldsymbol{d}^{t-1},d_{t}',\boldsymbol{U}^{t-1}(\boldsymbol{d}^{t-1}))
\end{array}\right]
\end{align*}
for $t\ge2$, by Assumption SX. Then, by Assumption RS and the discussion
above, this quantity is shown to be equal to
\begin{align}
 & \Pr\left[\left.\begin{array}{c}
\\
\boldsymbol{U}(d_{t}',\boldsymbol{d}_{-t})\in\mathcal{U}^{T}(d_{t}',\boldsymbol{d}_{-t},1;x_{t}',\boldsymbol{x}_{-t})\\
\\
\end{array}\right|\begin{array}{c}
\boldsymbol{U}^{t-1}(\boldsymbol{d}^{t-1})\in\mathcal{U}^{t-1}(\boldsymbol{d}^{t-1},\boldsymbol{y}^{t-1}),\\
\boldsymbol{V}^{t}\in\mathcal{V}^{t}(\boldsymbol{d}^{t-1},d_{t}',\boldsymbol{U}^{t-1}(\boldsymbol{d}^{t-1}))
\end{array}\right]\nonumber \\
= & E[Y_{T}(d_{t}',\boldsymbol{d}_{-t})|x_{t}',\boldsymbol{x}_{-t},\boldsymbol{z},\boldsymbol{y}^{t-1},\boldsymbol{d}^{t-1},d_{t}'],\label{eq:pf_thm1-2}
\end{align}
by Assumption SX. Note that this last quantity is still unobserved,
since $d_{s}$ for $s\ge t+1$ are not realized treatments; e.g.,
when $T=3$ and $t=2$,
\begin{align*}
E[Y_{3}(\boldsymbol{d})|\boldsymbol{x},\boldsymbol{z},y_{1},d_{1},d_{2}'] & =E[Y_{3}(d_{1},d_{2}',d_{3})|x_{1},x_{2}',x_{3},\boldsymbol{z},y_{1},d_{1},d_{2}'].
\end{align*}
The quantity, however, will be useful in the remaining proof where
we use mathematical induction to recover $E[Y_{T}(\boldsymbol{d})|\boldsymbol{x},\boldsymbol{z}]$;
see the Appendix. Recall the abbreviations $\mathcal{V}^{t}(\boldsymbol{d}^{t-1},d_{t}',\boldsymbol{U}^{t-1}(\boldsymbol{d}^{t-1}))\equiv\mathcal{V}^{t}(\boldsymbol{d}^{t-1},d_{t}',\boldsymbol{U}^{t-1}(\boldsymbol{d}^{t-1});\boldsymbol{z}^{t},\boldsymbol{x}^{t-1})$
and $\mathcal{U}^{t-1}(\boldsymbol{d}^{t-1},\boldsymbol{y}^{t-1})$
$\equiv\mathcal{U}^{t-1}(\boldsymbol{d}^{t-1},\boldsymbol{y}^{t-1};\boldsymbol{x}^{t-1})$.
That is, in the derivation of \eqref{eq:pf_thm1-2}, the key is to
consider the average potential outcome for a group of individuals
that is defined by the treatments at time $t$ or earlier and the
lagged outcome, for which $x_{t}$ is excluded.

The proof of Theorem \ref{thm:ASF} is constructive in that it provides
a closed-form expression for $E[Y_{T}(\boldsymbol{d})|\boldsymbol{x}]$
in an iterative manner, which can immediately be used for estimation.
For concreteness, we provide an expression for $E[Y_{T}(\boldsymbol{d})|\boldsymbol{x}]$
when $T=2$ and binary $Z_{t}$. Define
\begin{align*}
 & h_{t}^{d_{t}}(x_{t};\boldsymbol{y}^{t-1})\equiv h_{t}^{d_{t}}(x_{t};\boldsymbol{z}^{t-1},\boldsymbol{x}^{t-1},\boldsymbol{d}^{t-1},\boldsymbol{y}^{t-1})\\
\equiv & \Pr[Y_{t}=1,D_{t}=d_{t}|Z_{t}=1,\boldsymbol{z}^{t-1},\boldsymbol{x}^{t},\boldsymbol{d}^{t-1},\boldsymbol{y}^{t-1}]-\Pr[Y_{t}=1,D_{t}=d_{t}|Z_{t}=0,\boldsymbol{z}^{t-1},\boldsymbol{x}^{t},\boldsymbol{d}^{t-1},\boldsymbol{y}^{t-1}]
\end{align*}
and
\begin{align*}
\lambda_{t}(x_{t};\boldsymbol{y}^{t-1}) & \equiv\{\tilde{x}_{t}:h_{t}^{d_{t}}(x_{t};\boldsymbol{y}^{t-1})+h_{t}^{d_{t}'}(\tilde{x}_{t};\boldsymbol{y}^{t-1})=0\}
\end{align*}
with $\lambda_{1}(x_{1})\equiv\lambda_{1}(x_{1};y_{0})$. By Lemma
\ref{lem:sign_of_period_ATE}, $x_{t}'$ satisfies $\mu_{t}(\boldsymbol{y}^{t-1},\boldsymbol{d}^{t},x_{t})=\mu_{t}(\boldsymbol{y}^{t-1},\boldsymbol{d}^{t-1},d_{t}',x_{t}')$
if and only if $x_{t}'\in\lambda_{t}(x_{t};\boldsymbol{y}^{t-1})$.
Then, our identification result suggests that
\begin{align}
E[Y_{2}(\boldsymbol{d})|\boldsymbol{x}]= & \int\{P[\boldsymbol{d}|\boldsymbol{x},\boldsymbol{z}]E[Y_{2}|\boldsymbol{x},\boldsymbol{z},\boldsymbol{d}]+P[d_{1},d_{2}'|\boldsymbol{x},\boldsymbol{z}]\mu_{2,d_{1},d_{2}'}\nonumber \\
 & +P[d_{1}',d_{2}|\boldsymbol{x},\boldsymbol{z}]E[Y_{2}|\lambda_{1}(x_{1}),x_{2},\boldsymbol{z},d_{1}',d_{2}]+P[d_{1}',d_{2}'|\boldsymbol{x},\boldsymbol{z}]\mu_{2,d_{1}',d_{2}'}\}dF_{\boldsymbol{Z}|\boldsymbol{x}},\label{eq:ex_ID}
\end{align}
where
\begin{align*}
\mu_{2,d_{1},d_{2}'}\equiv & P[y_{1}|\boldsymbol{x},\boldsymbol{z},d_{1},d_{2}']E[Y_{2}|x_{1},\lambda_{2}(x_{2};y_{1}),\boldsymbol{z},d_{1},d_{2}',y_{1}]\\
 & +P[y_{1}'|\boldsymbol{x},\boldsymbol{z},d_{1},d_{2}']E[Y_{2}|x_{1},\lambda_{2}(x_{2};y_{1}'),\boldsymbol{z},d_{1},d_{2}',y_{1}'],\\
\mu_{2,d_{1}',d_{2}'}\equiv & P[y_{1}|\boldsymbol{x},\boldsymbol{z},d_{1}',d_{2}']E[Y_{2}|\lambda_{1}(x_{1}),\lambda_{2}(x_{2};y_{1}),\boldsymbol{z},d_{1}',d_{2}',y_{1}]\\
 & +P[y_{1}'|\boldsymbol{x},\boldsymbol{z},d_{1}',d_{2}']E[Y_{2}|\lambda_{1}(x_{1}),\lambda_{2}(x_{2};y_{1}'),\boldsymbol{z},d_{1}',d_{2}',y_{1}'].
\end{align*}
The aggregation with respect to $\boldsymbol{Z}=\boldsymbol{z}$ conditional
on $\boldsymbol{X}=\boldsymbol{x}$ is to improve efficiency. In Appendix
\ref{sec:Estimation-and-Inference}, we discuss further estimation
and inference strategies for the parameter $E[Y_{2}(\boldsymbol{d})|\boldsymbol{x}]$.

\begin{remark}\label{rem:initial_condi}The assumption that the initial
condition $Y_{0}$ is exogenously determined is not necessary but
imposed for convenience. Such an assumption appears in, e.g., \citet{heckman2007dynamic}.
In an alternative setting where $Y_{0}$ is endogenously determined
in the model, a similar identification analysis as in this section
can be followed by modifying Assumption SX. We may consider two alternatives
depending upon whether $Y_{0}$ is observable or not: (a) $(\boldsymbol{U}(\boldsymbol{d}),\boldsymbol{V})$
and $(\boldsymbol{Z},\boldsymbol{X})$ are independent conditional
on $Y_{0}$; or (b) $(\boldsymbol{U}(\boldsymbol{d}),\boldsymbol{V},Y_{0})$
and $(\boldsymbol{Z},\boldsymbol{X})$ are independent. First, recall
that each of these statements is ``conditional on other covariates.''
The assumption (a) can be imposed when $Y_{0}$ is observable, maybe
because $t=1$ is not the start of sample period. The assumption (b)
can be imposed when $Y_{0}$ is unobservable, maybe because $t=1$
is the start of sample period and the logical start of the process.
The analysis in these alternative scenarios is omitted as it is a
straightforward extension of the current one. In this analysis, there
is no need to assume the distribution of initial conditions, unlike
in the literature on dynamic models with random effects. Still, we
recover certain treatment effects, unlike in the literature on nonseparable
models with unobservable individual effects where, in general, partial
effects are hard to recover. The trade-off is that we require variables
that are independent of the individual effects, even though other
covariates are allowed not to be.\end{remark}

\begin{remark}\label{rem:about_SX}The strict exogeneity of Assumption
SX is a simple sufficient condition for what we actually need for
the identification analysis. As described in Lemma \ref{lem:high_level}
of the Appendix, the conditions we need to show Lemma \ref{lem:sign_of_period_ATE}
and Theorem \ref{thm:ASF}, respectively, are the following: For each
$t$, (i) $(Z_{t},X_{t})\perp(U_{t}(d_{t}),V_{t})|\boldsymbol{Z}^{t-1},\boldsymbol{X}^{t-1}$;
(ii) $Z_{t}\perp(\boldsymbol{U}(\boldsymbol{d}),\boldsymbol{V}^{t})|\boldsymbol{Z}^{t-1},\boldsymbol{X}_{-t}$
and $X_{t}\perp(\boldsymbol{U}(\boldsymbol{d}),\boldsymbol{V}^{t})|\boldsymbol{Z}^{t-1},\boldsymbol{X}_{-t}$.
In these high-level conditions, the condition for $Z_{t}$ is reminiscent
of the sequential randomization assumption. In fact, this is consistent
with our leading example of experimental studies with partial compliance.

\end{remark}

\begin{remark}\label{rem:rank_with_x}In order to define the $\boldsymbol{U}$-set,
recall that we use an alternative potential outcome $Y_{t}(\boldsymbol{d}^{t},\boldsymbol{x}^{t})=\mu_{t}(\boldsymbol{Y}^{t-1}(\boldsymbol{d}^{t-1},\boldsymbol{x}^{t-1}),\boldsymbol{d}^{t},x_{t},U_{t}(d_{t}))$.
Motivated from this, we may consider a structural model that adds
another dimension for heterogeneity by allowing $U_{t}(d_{t})$ to
be a function of $x_{t}$ as well:
\begin{align*}
Y_{t}(\boldsymbol{d}^{t},\boldsymbol{x}^{t}) & =\mu_{t}(\boldsymbol{Y}^{t-1}(\boldsymbol{d}^{t-1},\boldsymbol{x}^{t-1}),\boldsymbol{d}^{t},x_{t},U_{t}(d_{t},x_{t})).
\end{align*}
Given this extension, we can relax Assumption RS and impose that $\{\boldsymbol{U}(d_{t},\boldsymbol{d}_{-t},x_{t},\boldsymbol{x}_{-t})\}_{d_{t},x_{t}}$
are identically distributed conditional on $\boldsymbol{V}^{t}$ and
$(\boldsymbol{Z},\boldsymbol{X})$. The current Assumption RS can
be viewed as requiring rank invariance in terms of $x_{t}$, while
it allows rank similarity in $d_{t}$.

\end{remark}

\section{Treatment Effects on Transitions\label{sec:Treatment-Effects-on}}

In fact, the identification strategy introduced in the previous section
can tackle a more general problem. In this section, we extend the
identification analysis of the ATE (Theorem \ref{thm:ASF} and Corollary
\ref{cor:ID}) and show identification of the transition-specific
ATE. Given the vector $\boldsymbol{Y}(\boldsymbol{d})\equiv(Y_{1}(\boldsymbol{d}),...,Y_{T}(\boldsymbol{d}))$
of potential outcomes, let $\boldsymbol{Y}_{-}(\boldsymbol{d})\equiv(Y_{t_{1}}(\boldsymbol{d}),...,Y_{t_{L}}(\boldsymbol{d}))\in\mathcal{Y}_{-}\subseteq\{0,1\}^{L}$
be its $1\times L$ subvector, where $t_{1}<t_{2}<\cdots<t_{L}\le T-1$
and $L<T$. Then, the transition-specific ATE can be defined as $E[Y_{T}(\boldsymbol{d})|\boldsymbol{Y}_{-}(\boldsymbol{d})=\boldsymbol{y}_{-},\boldsymbol{X}=\boldsymbol{x}]-E[Y_{T}(\tilde{\boldsymbol{d}})|\boldsymbol{Y}_{-}(\tilde{\boldsymbol{d}})=\boldsymbol{y}_{-},\boldsymbol{X}=\boldsymbol{x}]$
for some sequences $\boldsymbol{d}$ and $\tilde{\boldsymbol{d}}$.
\begin{theorem}\label{thm:TSATE}Under Assumptions C, SX, RS and
SP, for each $\boldsymbol{y}_{-}$, $E[Y_{T}(\boldsymbol{d})|\boldsymbol{Y}_{-}(\boldsymbol{d})=\boldsymbol{y}_{-},\boldsymbol{X}=\boldsymbol{x}]-E[Y_{T}(\tilde{\boldsymbol{d}})|\boldsymbol{Y}_{-}(\tilde{\boldsymbol{d}})=\boldsymbol{y}_{-},\boldsymbol{X}=\boldsymbol{x}]$
is identified for $\boldsymbol{d},\tilde{\boldsymbol{d}}\in\mathcal{D}$
and $\boldsymbol{x}\in\mathcal{X}(\boldsymbol{d})\cap\mathcal{X}(\tilde{\boldsymbol{d}})$.\end{theorem}
The proof of this theorem extends that of Theorem \ref{thm:ASF};
see the Appendix.\footnote{As before, the parameters in Theorem \ref{thm:TSATE} and Corollary
\ref{cor:TSATE} below can be defined for any given period instead
of the terminal period $T$. The identification analysis of such parameters
is essentially the same, and thus omitted.} The transition-specific ATE defined in Theorem \ref{thm:TSATE} concerns
a transition from a state that is specified by the value of the vector
of previous potential outcomes, $\boldsymbol{Y}_{-}(\boldsymbol{d})$.
When $Y_{T}(\boldsymbol{d})$ is binary, $E[Y_{T}(\boldsymbol{d})|\boldsymbol{Y}_{-}(\boldsymbol{d})=\boldsymbol{y}_{-},\boldsymbol{X}=\boldsymbol{x}]$
can be viewed as a generalization of the transition probability. As
a simple example, with $L=T-1$, one may be interested in a transition
to one state when all previous potential outcomes have stayed in the
other state until $T-1$. When $L=1$ with $\boldsymbol{Y}_{-}(\boldsymbol{d})=Y_{T-1}(\boldsymbol{d})$,
the transition-specific ATE becomes $\Pr[Y_{T}(\boldsymbol{d})=1|Y_{T-1}(\boldsymbol{d})=0]-\Pr[Y_{T}(\tilde{\boldsymbol{d}})=1|Y_{T-1}(\tilde{\boldsymbol{d}})=0]$
introduced in Section \ref{sec:model_and_para}. This is a particular
example of the treatment effect on the transition probability. The
treatment effects on transitions have been studied by, e.g., \citet{abbring2003nonparametric},
\citet{heckman2007dynamic}, \citet{fredriksson2008dynamic} and \citet{vikstrom2018bounds}.\footnote{The definition of the treatment effect on the transition probability
in this paper differs from those defined in the literature on duration
models, e.g., that in \citet{vikstrom2018bounds}. Since \citet{vikstrom2018bounds}'s
main focus is on $Y_{t}$ that is irreversible, they define a different
treatment parameter that yields a specific interpretation under dynamic
selection; see their paper for details. In addition, they assume sequential
randomization and that treatments are assigned earlier than the transition
of interest.} Let $Y_{t}(d_{t})\equiv\mu_{t}(Y_{t-1},d_{t},X_{t},U_{t}(d_{t}))$
be the \textit{period-specific potential outcome} at time $t$. Since
$Y_{t-1}=Y_{t-1}(\boldsymbol{D}^{t-1})$, the period-specific potential
outcome can be expressed as $Y_{t}(d_{t})=Y_{t}(\boldsymbol{D}^{t-1},d_{t})$
using the usual potential outcome. As a corollary of the result above,
we also identify a related parameter that specifies the previous state
by the observed outcome: $E[Y_{T}(1)-Y_{T}(0)|Y_{T-1}=y_{T-1}]$.\begin{corollary}\label{cor:TSATE}Under
Assumptions C, SX, RS and SP, for each $y_{T-1}$, $E[Y_{T}(1)|y_{T-1},\boldsymbol{x}]-E[Y_{T}(0)|y_{T-1},\boldsymbol{x}]$
is identified for $\boldsymbol{x}\in\mathcal{X}(\boldsymbol{d})\cap\mathcal{X}(\tilde{\boldsymbol{d}})$.\end{corollary}The
corollary is derived by observing that $Y_{T}(d_{T})=Y_{T}(\boldsymbol{D}^{T-1},d_{T})$,
and thus
\begin{align*}
 & E[Y_{T}(d_{T})|y_{T-1},\boldsymbol{x}]\\
= & \sum_{\boldsymbol{d}^{T-1}\in\mathcal{D}^{T-1}}\Pr[\boldsymbol{D}^{T-1}=\boldsymbol{d}^{T-1}|\boldsymbol{x}]E[Y_{T}(\boldsymbol{d}^{T-1},d_{T})|Y_{T-1}(\boldsymbol{d}^{T-1})=y_{T-1},\boldsymbol{D}^{T-1}=\boldsymbol{d}^{T-1},\boldsymbol{x}],
\end{align*}
where each $E[Y_{T}(\boldsymbol{d}^{T-1},d_{T})|Y_{T-1}(\boldsymbol{d}^{T-1})=y_{T-1},\boldsymbol{d}^{T-1},\boldsymbol{x}]$
is identified from the iteration at $t=T-1$ in the proof of Theorem
\ref{thm:TSATE} by taking $Y_{-}(\boldsymbol{d})=Y_{T-1}(\boldsymbol{d}^{T-1})$.

\section{Partial Identification\label{sec:partial_ID}}

Suppose Assumption SP does not hold in that $X_{t}$ does not exhibit
sufficient rectangular variation, or that there is no $X_{t}$ that
is excluded from the selection equation at time $t$. In this case,
we partially identify the ARSF's, ATE's and $\boldsymbol{d}^{*}(\boldsymbol{x})$
(or $\boldsymbol{d}^{\dagger}(\boldsymbol{x})$).

We briefly illustrate the calculation of the bounds on the ARSF $E[Y_{T}(\boldsymbol{d})|\boldsymbol{x}]$
when the sufficient rectangular variation is not guaranteed; the case
where $X_{t}$ does not exist at all can be dealt in a similar manner,
and so is omitted. For each $E[Y_{T}(\boldsymbol{d})|\boldsymbol{x},\boldsymbol{z},\boldsymbol{y}^{t-1},\boldsymbol{d}^{t-1},d_{t}']$
in the proof of Theorem \ref{thm:ASF}, we can calculate its upper
and lower bounds depending on the sign of $\mu_{t}(\boldsymbol{y}^{t-1},\boldsymbol{d}^{t-1},1,x_{t})-\mu_{t}(\boldsymbol{y}^{t-1},\boldsymbol{d}^{t-1},0,\tilde{x}_{t})$,
which is identified in Lemma \ref{lem:sign_of_period_ATE}. Note that,
in the context of this section, $\tilde{x}_{t}$ does \textit{not}
necessarily differ from $x_{t}$. For example, for the lower bound
on $E[Y_{T}(\boldsymbol{d})|\boldsymbol{x}]=E[Y_{T}(\boldsymbol{d})|\boldsymbol{x},\boldsymbol{z}]$,
suppose $\mu_{t}(\boldsymbol{y}^{t-1},\boldsymbol{d}^{t},x_{t})-\mu_{t}(\boldsymbol{y}^{t-1},\boldsymbol{d}^{t-1},d_{t}',x_{t}')\ge0$
for given $(\boldsymbol{y}^{t-1},\boldsymbol{d}^{t-1})$, where $x_{t}'$
is allowed to equal $x_{t}$. Then, by the definition of the $\boldsymbol{U}$-set
and under Assumption RS, it satisfies that $\mathcal{U}^{T}(\boldsymbol{d},y_{T};\boldsymbol{x})\supseteq\mathcal{U}^{T}(d_{t}',\boldsymbol{d}_{-t},y_{T};x_{t}',\boldsymbol{x}_{-t})$,
conditional on $\boldsymbol{Y}^{t-1}(\boldsymbol{d}^{t-1},\boldsymbol{x}^{t-1})=\boldsymbol{y}^{t-1}$.
Therefore, we have a lower bound on as $E[Y_{T}(\boldsymbol{d})|\boldsymbol{x},\boldsymbol{z},\boldsymbol{y}^{t-1},\boldsymbol{d}^{t-1},d_{t}']$
as
\begin{align}
 & E[Y_{T}(\boldsymbol{d})|\boldsymbol{x},\boldsymbol{z},\boldsymbol{y}^{t-1},\boldsymbol{d}^{t-1},d_{t}']\nonumber \\
= & \Pr[\boldsymbol{U}(\boldsymbol{d})\in\mathcal{U}^{T}(\boldsymbol{d},1;\boldsymbol{x})|\boldsymbol{U}^{t-1}(\boldsymbol{d}^{t-1})\in\mathcal{U}^{t-1}(\boldsymbol{d}^{t-1},\boldsymbol{y}^{t-1}),\boldsymbol{V}^{t}\in\mathcal{V}^{t}(\boldsymbol{d}^{t-1},d_{t}',\boldsymbol{U}^{t-1}(\boldsymbol{d}^{t-1}))]\nonumber \\
\ge & \Pr\left[\left.\begin{array}{c}
\\
\boldsymbol{U}(d_{t}',\boldsymbol{d}_{-t})\in\mathcal{U}^{T}(d_{t}',\boldsymbol{d}_{-t},1;x_{t}',\boldsymbol{x}_{-t})\\
\\
\end{array}\right|\begin{array}{c}
\boldsymbol{U}^{t-1}(\boldsymbol{d}^{t-1})\in\mathcal{U}^{t-1}(\boldsymbol{d}^{t-1},\boldsymbol{y}^{t-1}),\\
\boldsymbol{V}^{t}\in\mathcal{V}^{t}(\boldsymbol{d}^{t-1},d_{t}',\boldsymbol{U}^{t-1}(\boldsymbol{d}^{t-1}))
\end{array}\right]\nonumber \\
= & E[Y_{T}(d_{t}',\boldsymbol{d}_{-t})|x_{t}',\boldsymbol{x}_{-t},\boldsymbol{z},\boldsymbol{y}^{t-1},\boldsymbol{d}^{t-1},d_{t}'].\label{eq:lower_bound}
\end{align}
Then, it is possible to calculate the lower bounds on $E[Y_{T}(\boldsymbol{d})|\boldsymbol{x},\boldsymbol{z}]$
using the iterative scheme introduced in the proof of Theorem \ref{thm:ASF}.
That is, at each iteration, we take the previous iteration's lower
bound as given, expand each main term in \eqref{eq:expand} as before,
and apply \eqref{eq:lower_bound} for necessary terms.

Lastly, depending on the signs of the ATE's, we can construct bounds
on $\boldsymbol{d}^{*}(w_{0})$ (or $\boldsymbol{d}^{\dagger}(w_{0})$),
which will be expressed as strict subsets of $\mathcal{D}$. The
partial identification of the optimal regimes may not yield sufficiently
narrow bounds unless there are a sufficient number of ATE's whose
bounds are informative about their signs. In general, however, the
informativeness of bounds truly depends on the policy questions. Note
that $\mathcal{D}$ is a discrete set. Even though the bounds may
not be informative about the optimal regime, they may still be useful
from the planner's perspective if they can help her exclude a few
suboptimal regimes, i.e., $\boldsymbol{d}^{\circ}$ such that $E[Y_{T}(\boldsymbol{d})|w_{0}]\ge E[Y_{T}(\boldsymbol{d}^{\circ})|w_{0}]$
for some $\boldsymbol{d}$.

\section{Subsequences of Treatments\label{sec:subseq}}

An important extension of the model introduced in this paper is to
the case where treatments do not appear in every period, while the
outcomes are constantly observed. For example, institutionally, there
may only be a one-shot treatment at the beginning of time or a few
treatments earlier in the horizon, or there may be evenly spaced treatment
decisions with a lower frequency than outcomes. A potential outcome
that corresponds to this situation can be defined as a function of
a certain subsequence $\boldsymbol{d}_{-}$ of $\boldsymbol{d}$.
Let $\boldsymbol{d}_{-}\equiv(d_{t_{1}},...,d_{t_{K}})\in\mathcal{D}_{-}\subseteq\{0,1\}^{K}$
be a $1\times K$ subvector of $\boldsymbol{d}$, where $t_{1}<t_{2}<\cdots<t_{K}\le T$
and $K<T$. Then, the potential outcomes $Y_{t}(\boldsymbol{d}_{-})$
and the associated structural functions are defined as follows: Let
$\boldsymbol{d}_{-}^{t_{k}}\equiv(d_{t_{1}},...,d_{t_{k}})$. A potential
outcome in the period when a treatment exists is expressed using a
switching regression model as
\begin{align*}
Y_{t_{k}}(\boldsymbol{d}_{-}) & =Y_{t_{k}}(\boldsymbol{d}_{-}^{t_{k}})=\mu_{t_{k}}(\boldsymbol{Y}^{t_{k}-1}(\boldsymbol{d}_{-}^{t_{(k-1)}}),\boldsymbol{d}^{t_{k}},X_{t_{k}},U_{t_{k}}(d_{t_{k}}))
\end{align*}
for $k\ge1$ with $Y_{t_{1}-1}(\boldsymbol{d}_{-}^{t_{0}})=Y_{t_{1}-1}$,
and a potential outcome when there is no treatment is expressed as
\begin{align*}
Y_{t}(\boldsymbol{d}_{-}) & =Y_{t}(\boldsymbol{d}_{-}^{t_{k}})=\mu_{t}(\boldsymbol{Y}^{t-1}(\boldsymbol{d}_{-}^{t_{k}}),U_{t})
\end{align*}
for $t$ such that $t_{k}<t<t_{(k+1)}$ ($1\le k\le K-1$). Lastly,
$Y_{t}(\boldsymbol{d}_{-})=Y_{t}=\mu_{t}(\boldsymbol{Y}^{t-1},U_{t})$
for $t<t_{1}$ and $Y_{t}(\boldsymbol{d}_{-})=Y_{t}(\boldsymbol{d}_{-}^{t_{K}})=\mu_{t}(\boldsymbol{Y}^{t-1}(\boldsymbol{d}_{-}^{t_{K}}),U_{t})$
for $t>t_{K}$. Each structural model at the time of no treatment
is a plain dynamic model with a lagged dependent variable. Let $T=4$
and $\boldsymbol{d}_{-}=(d_{1},d_{3})$ for illustration. Then the
sequence of potential outcomes can be expressed as
\begin{align*}
Y_{4}(\boldsymbol{d}_{-}) & =Y_{4}(\boldsymbol{d}^{3})=\mu_{4}(\boldsymbol{Y}^{3}(\boldsymbol{d}^{3}),U_{4}),\\
Y_{3}(\boldsymbol{d}_{-}) & =Y_{3}(\boldsymbol{d}^{3})=\mu_{3}(\boldsymbol{Y}^{2}(d_{1}),\boldsymbol{d}^{3},X_{3},U_{3}(d_{3})),\\
Y_{2}(\boldsymbol{d}_{-}) & =Y_{2}(d_{1})=\mu_{2}(Y_{1}(d_{1}),U_{2}),\\
Y_{1}(\boldsymbol{d}_{-}) & =Y_{1}(d_{1})=\mu_{1}(Y_{0},d_{1},X_{1},U_{1}(d_{1})).
\end{align*}
The selection equations are of the following form: For $k\ge1$,
\begin{align*}
D_{t_{k}} & =1\{\pi_{t_{k}}(\boldsymbol{Y}^{t_{k}-1},\boldsymbol{D}^{t_{(k-1)}},Z_{t_{k}})\ge V_{t_{k}}\},
\end{align*}
where the lagged outcome and the \textit{latest} treatment enter each
equation. The observable variables are $(\boldsymbol{Y},\boldsymbol{D}_{-},\boldsymbol{X}_{-},\boldsymbol{Z}_{-})$.\footnote{It may be the case that $X_{t}$ is observed whenever $Y_{t}$ is
observed, and thus is included in the $Y_{t}$-equations for $t\neq t_{k}$
as well. We ignore that case here.}

Now all the parameters introduced in Section \ref{sec:model_and_para}
can be readily modified by replacing $\boldsymbol{d}$ with $\boldsymbol{d}_{-}$
for some $\boldsymbol{d}_{-}$; we omit the definitions for the sake
of brevity. Moreover, the identification analysis of Section \ref{sec:ID}
can be easily modified in accordance with the extended setting. Let
$\boldsymbol{U}_{-}(\boldsymbol{d}_{-})\equiv(U_{t_{1}}(d_{t_{1}}),...,U_{t_{K}}(d_{t_{K}}))$
and let $\boldsymbol{U}(\boldsymbol{d}_{-})$ be the vector of all
the outcome unobservables that consists of $\boldsymbol{U}_{-}(\boldsymbol{d}_{-})$
and $\{U_{t}\}_{t\in\{1,...,T\}\backslash\{t_{1},...,t_{K}\}}$.\begin{asC2}The
distribution of $(\boldsymbol{U}_{-}(\boldsymbol{d}_{-}),\boldsymbol{V}_{-})$
has strictly positive density with respect to Lebesgue measure on
$\mathbb{R}^{2K}$.\end{asC2}\begin{asSX2}$(\boldsymbol{Z}_{-},\boldsymbol{X}_{-})$
and $(\boldsymbol{U}(\boldsymbol{d}_{-}),\boldsymbol{V}_{-})$ are
independent.\end{asSX2}Let $\boldsymbol{d}_{-,-t_{k}}$ be $\boldsymbol{d}_{-}$
without the $t_{k}$-th element.\begin{asRS2}For each $t_{k}$ and
$\boldsymbol{d}_{-,-t_{k}}$, $\{\boldsymbol{U}_{-}(d_{t_{k}},\boldsymbol{d}_{-,-t_{k}})\}_{d_{t_{k}}}$
are identically distributed conditional on $(\boldsymbol{U}^{t_{k}-1}(\boldsymbol{d}_{-}^{t_{(k-1)}}),\boldsymbol{V}_{-}^{t_{k}})$.\end{asRS2}Under
these modified assumptions, Lemma \ref{lem:sign_of_period_ATE} is
now only relevant for $t=t_{k}$. Restrict the definitions of $\mathcal{X}_{t}(d_{t};\boldsymbol{x}_{-t},\boldsymbol{z}_{-t})$
in \eqref{eq:set4} and $\mathcal{X}_{t}(d_{t};\boldsymbol{x}_{-t})$
to hold only for $t=t_{k}$.\begin{asSP2}For each $t_{k}$ and $d_{t_{k}}$,
$\Pr[X_{t_{k}}\in\mathcal{X}_{t_{k}}(d_{t_{k}};\boldsymbol{x}_{-,-t_{k}},\boldsymbol{z}_{-,-t_{k}})|\boldsymbol{x}_{-,-t_{k}},\boldsymbol{z}_{-,-t_{k}}]>0$
almost everywhere.\end{asSP2}Let $\mathcal{X}_{-}(\boldsymbol{d}_{-})\equiv\{\boldsymbol{x}_{-}:x_{t_{k}}\in\mathcal{X}_{t_{k}}(d_{t_{k}};\boldsymbol{x}_{-,-t_{k}})\text{ for some }(x_{t_{(k+1)}},...,x_{t_{K}}),\text{ for }k\ge1\}$.\begin{theorem}\label{thm:ASF2}Under
Assumptions C$^{\prime}$, SX$^{\prime}$, RS$^{\prime}$ and SP$^{\prime}$,
$E[Y_{T}(\boldsymbol{d}_{-})|\boldsymbol{x}_{-}]$ is identified for
$\boldsymbol{d}_{-}\in\mathcal{D}_{-}$, $\boldsymbol{x}_{-}\in\mathcal{X}_{-}(\boldsymbol{d}_{-})$.\end{theorem}\begin{corollary}\label{cor:ID2}Under
Assumptions C$^{\prime}$, SX$^{\prime}$, RS$^{\prime}$ and SP$^{\prime}$,
$E[Y_{T}(\boldsymbol{d}_{-})-Y_{T}(\tilde{\boldsymbol{d}}_{-})|\boldsymbol{x}_{-}]$
is identified for $\boldsymbol{d}_{-},\tilde{\boldsymbol{d}}_{-}\in\mathcal{D}_{-}$
and $\boldsymbol{x}_{-}\in\mathcal{X}_{-}(\boldsymbol{d}_{-})\cap\mathcal{X}_{-}(\tilde{\boldsymbol{d}}_{-})$,
and $\boldsymbol{d}_{-}^{*}(w_{0})$ and $\boldsymbol{d}_{-}^{\dagger}(w_{0})$
are identified for $w_{0}\in\mathcal{W}_{0}$.\end{corollary}

\section{Conclusions\label{sec:Conclusions}}

In this paper, we consider identification in a nonparametric model
for dynamic treatments and outcomes. We introduce a sequence of selection
models, replacing the assumption of sequential randomization, which
may be hard to justify under partial compliance or in observational
settings. We consider treatment and outcome processes of general forms,
and avoid making strong assumptions on distribution and functional
forms, nor assumptions on rationality. We show that the treatment
parameters and optimal treatment regimes are point identified under
the two-way exclusion restriction and sequential rank similarity.
We argue that the reverse exclusion restriction is a useful alternative
tool for empirical researchers who seek identification in this type
of nonseparable models with endogeneity. This source of variation
may especially be easy to find and justify in a dynamic setting as
in this paper. When the reverse exclusion restriction is violated,
we show how to characterize bounds on these parameters.

\medskip{}

\bibliographystyle{ecta}
\bibliography{dynTE}

\pagebreak{}

\begin{appendix}

\section{Estimation and Inference\label{sec:Estimation-and-Inference}}

The identification analysis is constructive and naturally suggests
an estimation procedure by the sample analog principle. Here we illustrate
that by considering \eqref{eq:ex_ID}, which can be alternatively
expressed as
\begin{align*}
E[Y_{2}(\boldsymbol{d})|\boldsymbol{x}]=\int\{ & E[1_{\boldsymbol{d}}Y_{2}|\boldsymbol{x},\boldsymbol{z}]+E[1_{d_{1},d_{2}'}1_{y_{1}}Y_{2}|x_{1},\lambda_{2}(x_{2};y_{1}),\boldsymbol{z}]+E[1_{d_{1},d_{2}'}1_{y_{1}'}Y_{2}|x_{1},\lambda_{2}(x_{2};y_{1}'),\boldsymbol{z}]\\
 & +E[1_{d_{1}',d_{2}}Y_{2}|\lambda_{1}(x_{1}),x_{2},\boldsymbol{z}]\\
 & +E[1_{d_{1}',d_{2}'}1_{y_{1}}|x_{1},\boldsymbol{z}]E[Y_{2}|\lambda_{1}(x_{1}),\lambda_{2}(x_{2};y_{1}),\boldsymbol{z},d_{1}',d_{2}',y_{1}]\\
 & +E[1_{d_{1}',d_{2}'}1_{y_{1}'}|x_{1},\boldsymbol{z}]E[Y_{2}|\lambda_{1}(x_{1}),\lambda_{2}(x_{2};y_{1}'),\boldsymbol{z},d_{1}',d_{2}',y_{1}']\}dF_{\boldsymbol{Z}|\boldsymbol{x}},
\end{align*}
where $1_{\boldsymbol{d}}\equiv1[\boldsymbol{D}=\boldsymbol{d}]$
for generic $\boldsymbol{d}$, the second term on the right hand side
is by $E[1_{d_{1},d_{2}'}1_{y_{1}}|x_{1},\boldsymbol{z}]\times$ $E[Y_{2}|x_{1},\lambda_{2}(x_{2};y_{1}),\boldsymbol{z},d_{1},d_{2}',y_{1}]=E[1_{d_{1},d_{2}'}1_{y_{1}}Y_{2}|x_{1},\lambda_{2}(x_{2};y_{1}),\boldsymbol{z}]$
since $E[1_{d_{1},d_{2}'}1_{y_{1}}|\boldsymbol{x},\boldsymbol{z}]=E[1_{d_{1},d_{2}'}1_{y_{1}}|x_{1},\boldsymbol{z}]$
(and similarly for the third term), and the fourth term is by $P[d_{1}',d_{2}|\boldsymbol{x},\boldsymbol{z}]=P[d_{1}',d_{2}|\boldsymbol{z}]$.
When $\lambda_{1}(\cdot)$ and $\lambda_{2}(\cdot;\cdot)$ are assumed
to be known, the estimation of $g_{\boldsymbol{d}}(\boldsymbol{x})\equiv E[Y_{2}(\boldsymbol{d})|\boldsymbol{x}]$
and inference on its functionals can be dealt as a special case of
the penalized sieve minimum distance (PSMD) estimation framework of
\citet{chen2015sieve} by constructing the following conditional moments:
\begin{align*}
\sum_{\boldsymbol{z}\in\{0,1\}^{2}}\Pr[\boldsymbol{Z}_{i}=\boldsymbol{z}|\boldsymbol{x}]\{g_{\boldsymbol{z}}^{1}(\boldsymbol{x})+g_{\boldsymbol{z}}^{2}(x_{1})+g_{\boldsymbol{z}}^{3}(x_{1})\qquad\qquad\qquad\\
+g_{\boldsymbol{z}}^{4}(x_{2})+g_{\boldsymbol{z}}^{5}(x_{1})\mu_{\boldsymbol{z}}^{1}+g_{\boldsymbol{z}}^{6}(x_{1})\mu_{\boldsymbol{z}}^{2}\}-g_{\boldsymbol{d}}(\boldsymbol{x}) & =0,\\
E[1_{\boldsymbol{d}}Y_{2}|\boldsymbol{x},\boldsymbol{z}]-g_{\boldsymbol{z}}^{1}(\boldsymbol{x}) & =0,\\
E[1_{d_{1},d_{2}'}1_{y_{1}}Y_{2}|x_{1},\lambda_{2}(x_{2};y_{1}),\boldsymbol{z}]-g_{\boldsymbol{z}}^{2}(x_{1}) & =0,\\
E[1_{d_{1},d_{2}'}1_{y_{1}'}Y_{2}|x_{1},\lambda_{2}(x_{2};y_{1}'),\boldsymbol{z}]-g_{\boldsymbol{z}}^{3}(x_{1}) & =0,\\
E[1_{d_{1}',d_{2}}Y_{2}|\lambda_{1}(x_{1}),x_{2},\boldsymbol{z}]-g_{\boldsymbol{z}}^{4}(x_{2}) & =0,\\
E[1_{d_{1}',d_{2}'}1_{y_{1}}|x_{1},\boldsymbol{z}]-g_{\boldsymbol{z}}^{5}(x_{1}) & =0,\\
E[1_{d_{1}',d_{2}'}1_{y_{1}'}|x_{1},\boldsymbol{z}]-g_{\boldsymbol{z}}^{6}(x_{1}) & =0,\\
E[Y_{2}|\lambda_{1}(x_{1}),\lambda_{2}(x_{2};y_{1}),\boldsymbol{z},d_{1}',d_{2}',y_{1}]-\mu_{\boldsymbol{z}}^{1} & =0,\\
E[Y_{2}|\lambda_{1}(x_{1}),\lambda_{2}(x_{2};y_{1}'),\boldsymbol{z},d_{1}',d_{2}',y_{1}']-\mu_{\boldsymbol{z}}^{2} & =0,
\end{align*}
where $g_{\boldsymbol{d}}(\boldsymbol{x})$, $\Pr[\boldsymbol{Z}_{i}=\boldsymbol{z}|\cdot]$
and $g_{\boldsymbol{z}}^{j}(\cdot)$ for $j\in\{1,...,6\}$ are the
nonparametric components, and $\mu_{\boldsymbol{z}}^{1}$ and $\mu_{\boldsymbol{z}}^{2}$
are the parametric components, for $\boldsymbol{z}\in\{0,1\}^{2}$.
It is worth noting that, since none of the nonparametric components
has endogenous variables as its arguments, there is no ill-posed inverse
problem. For a (possibly nonlinear) functional $\phi(\cdot)$ of $g_{\boldsymbol{d}}(\boldsymbol{x})$,
the plug-in PSMD estimator $\phi(\hat{g}_{\boldsymbol{d}})$ is asymptotically
normal with a consistent sieve variance estimator, which can be use
to conduct inference. \citet{chen2015sieve} also establish asymptotic
theory for the sieve quasi likelihood ratio statistic, whose null
distribution is tight no matter whether $\phi(g_{\boldsymbol{d}})$
is $\sqrt{n}$-estimable (as with a weighted derivative functional)
or not (as with a point evaluation functional). Generalized weighted
bootstrap can be used to calculate critical values for this test statistic.
When the sets $\lambda_{1}(\cdot)$ and $\lambda_{2}(\cdot;\cdot)$
are estimated, estimation and inference become analogous to those
with the matching estimator in \citet{heckman1998matching}.

\section{Proofs\label{sec:Proofs}}

\subsection{High-Level Conditions for Assumption SX}

As discussed in Remark \ref{rem:about_SX}, Assumption SX is a sufficient
condition for high-level conditions for the proofs of Lemma \ref{lem:sign_of_period_ATE}
and Theorem \ref{thm:ASF}.

\begin{lemma}\label{lem:high_level}Assumption SX implies the following:
(i) $(Z_{t},X_{t})\perp(U_{t}(d_{t}),V_{t})|\boldsymbol{Z}^{t-1},\boldsymbol{X}^{t-1}$;
(ii) $Z_{t}\perp(\boldsymbol{U}(\boldsymbol{d}),\boldsymbol{V}^{t})|\boldsymbol{Z}^{t-1},\boldsymbol{X}_{-t}$
and $X_{t}\perp(\boldsymbol{U}(\boldsymbol{d}),\boldsymbol{V}^{t})|\boldsymbol{Z}^{t-1},\boldsymbol{X}_{-t}$.\end{lemma}

In the proofs below, we use these high-level conditions. Therefore,
some of the intermediate results we obtain in the proofs are slightly
different from the ones described in the main text for which Assumption
SX is directly applied.

\subsection{Proof of Lemma \ref{lem:sign_of_period_ATE}}

\noindent We first define the $\boldsymbol{U}$-set and $\boldsymbol{V}$-set.
The $\boldsymbol{U}$-set is defined in the main text. Realizing the
dependence of $Y_{s-1}(\boldsymbol{d}^{s-1},\boldsymbol{x}^{s-1})$
on $(\boldsymbol{U}^{s-1}(\boldsymbol{d}^{s-1}),\boldsymbol{x}^{s-1},\boldsymbol{d}^{s-1})$,
let
\begin{align*}
\pi_{s}^{*}(\boldsymbol{U}^{s-1}(\boldsymbol{d}^{s-1}),\boldsymbol{x}^{s-1},\boldsymbol{d}^{s-1},z_{s}) & \equiv\pi_{s}(Y_{s-1}(\boldsymbol{d}^{s-1},\boldsymbol{x}^{s-1}),d_{s-1},z_{s}),
\end{align*}
and define the set of $\boldsymbol{V}^{t}$ as
\begin{align*}
\mathcal{V}^{t}(\boldsymbol{d}^{t},\boldsymbol{u}^{t-1})\equiv\mathcal{V}^{t}(\boldsymbol{d}^{t},\boldsymbol{u}^{t-1};\boldsymbol{z}^{t},\boldsymbol{x}^{t-1}) & \equiv\{\boldsymbol{V}^{t}:d_{s}=1\{V_{s}\le\pi_{s}^{*}(\boldsymbol{u}^{s-1},\boldsymbol{x}^{s-1},\boldsymbol{d}^{s-1},z_{s})\}\text{ for all }s\le t\}
\end{align*}
for $t\ge2$. Fix $t\ge3$. Given \eqref{eq:rf_condi}, consider
the case $\Pr[D_{t}=1|\boldsymbol{z}^{t},\boldsymbol{x}^{t-1},\boldsymbol{d}^{t-1},\boldsymbol{y}^{t-1}]>\Pr[D_{t}=1|\tilde{z}_{t},\boldsymbol{z}^{t-1},\boldsymbol{x}^{t-1},\boldsymbol{d}^{t-1},\boldsymbol{y}^{t-1}]$;
the opposite case is symmetric. Using the definitions of the sets
above, we have
\begin{align*}
 & \Pr[D_{t}=1|\boldsymbol{z}^{t},\boldsymbol{x}^{t-1},\boldsymbol{d}^{t-1},\boldsymbol{y}^{t-1}]\\
= & \Pr[V_{t}\le\pi_{t}(\boldsymbol{y}^{t-1},\boldsymbol{d}^{t-1},z_{t})|\boldsymbol{z}^{t},\boldsymbol{x}^{t-1},\mathcal{V}^{t-1}(\boldsymbol{d}^{t-1},\boldsymbol{U}^{t-2}(\boldsymbol{d}^{t-2})),\mathcal{U}^{t-1}(\boldsymbol{d}^{t-1},\boldsymbol{y}^{t-1})]\\
= & \Pr[V_{t}\le\pi_{t}(\boldsymbol{y}^{t-1},\boldsymbol{d}^{t-1},z_{t})|\boldsymbol{z}^{t-1},\boldsymbol{x}^{t-1},\mathcal{V}^{t-1}(\boldsymbol{d}^{t-1},\boldsymbol{U}^{t-2}(\boldsymbol{d}^{t-2})),\mathcal{U}^{t-1}(\boldsymbol{d}^{t-1},\boldsymbol{y}^{t-1})],
\end{align*}
where the last equality is given by Assumption SX and Lemma \ref{lem:high_level}(i).
Note that the sets $\mathcal{V}^{t-1}(\boldsymbol{d}^{t-1},\boldsymbol{U}^{t-2}(\boldsymbol{d}^{t-2}))$
and $\mathcal{U}^{t-1}(\boldsymbol{d}^{t-1},\boldsymbol{y}^{t-1})$
do not change with the change in $z_{t}$. Therefore, a parallel expression
can be derived for $\Pr[D_{t}=1|\tilde{z}_{t},\boldsymbol{z}^{t-1},\boldsymbol{x}^{t-1},\boldsymbol{d}^{t-1},\boldsymbol{y}^{t-1}]$.
Let $\pi_{t}\equiv(\boldsymbol{y}^{t-1},\boldsymbol{d}^{t-1},z_{t})$
and $\tilde{\pi}_{t}\equiv(\boldsymbol{y}^{t-1},\boldsymbol{d}^{t-1},\tilde{z}_{t})$
for abbreviation. Then, under Assumption C,
\begin{align*}
0< & \Pr[D_{t}=1|\boldsymbol{z}^{t},\boldsymbol{x}^{t-1},\boldsymbol{d}^{t-1},\boldsymbol{y}^{t-1}]-\Pr[D_{t}=1|\tilde{z}_{t},\boldsymbol{z}^{t-1},\boldsymbol{x}^{t-1},\boldsymbol{d}^{t-1},\boldsymbol{y}^{t-1}]\\
= & \Pr[V_{t}\le\pi_{t}|\boldsymbol{z}^{t-1},\boldsymbol{x}^{t-1},\mathcal{V}^{t-1}(\boldsymbol{d}^{t-1},\boldsymbol{U}^{t-2}(\boldsymbol{d}^{t-2})),\mathcal{U}^{t-1}(\boldsymbol{d}^{t-1},\boldsymbol{y}^{t-1})]\\
 & -\Pr[V_{t}\le\tilde{\pi}_{t}|\boldsymbol{z}^{t-1},\boldsymbol{x}^{t-1},\mathcal{V}^{t-1}(\boldsymbol{d}^{t-1},\boldsymbol{U}^{t-2}(\boldsymbol{d}^{t-2})),\mathcal{U}^{t-1}(\boldsymbol{d}^{t-1},\boldsymbol{y}^{t-1})],
\end{align*}
which implies $\pi_{t}>\tilde{\pi}_{t}$. Next, we have
\begin{align*}
 & \Pr[Y_{t}=1,D_{t}=1|\boldsymbol{z}^{t},\boldsymbol{x}^{t},\boldsymbol{d}^{t-1},\boldsymbol{y}^{t-1}]\\
= & \Pr[U_{t}(1)\le\mu_{t}(\boldsymbol{y}^{t-1},\boldsymbol{d}^{t-1},1,x_{t}),V_{t}\le\pi_{t}|\boldsymbol{z}^{t-1},\boldsymbol{x}^{t-1},\mathcal{V}^{t-1}(\boldsymbol{d}^{t-1},\boldsymbol{U}^{t-2}(\boldsymbol{d}^{t-2})),\mathcal{U}^{t-1}(\boldsymbol{d}^{t-1},\boldsymbol{y}^{t-1})]
\end{align*}
by Assumption SX and Lemma \ref{lem:high_level}(i). Again, note that
$\mathcal{V}^{t-1}(\boldsymbol{d}^{t-1},\boldsymbol{U}^{t-2}(\boldsymbol{d}^{t-2}))$
and $\mathcal{U}^{t-1}(\boldsymbol{d}^{t-1},\boldsymbol{y}^{t-1})$
do not change with the change in $(z_{t},x_{t})$, which is key. Therefore,
similar expressions can be derived for the other terms involved in
$h_{t}$, and we have
\begin{align*}
 & h_{t}(z_{t},\tilde{z}_{t},x_{t},\tilde{x}_{t};\boldsymbol{z}^{t-1},\boldsymbol{x}^{t-1},\boldsymbol{d}^{t-1},\boldsymbol{y}^{t-1})\\
= & \Pr[U_{t}(1)\le\mu_{t}(\boldsymbol{y}^{t-1},\boldsymbol{d}^{t-1},1,x_{t}),\tilde{\pi}_{t}\le V_{t}\le\pi_{t}|\boldsymbol{z}^{t-1},\boldsymbol{x}^{t-1},\mathcal{V}^{t-1}(\boldsymbol{d}^{t-1},\boldsymbol{U}^{t-2}(\boldsymbol{d}^{t-2})),\mathcal{U}^{t-1}(\boldsymbol{d}^{t-1},\boldsymbol{y}^{t-1})]\\
 & -\Pr[U_{t}(0)\le\mu_{t}(\boldsymbol{y}^{t-1},\boldsymbol{d}^{t-1},0,\tilde{x}_{t}),\tilde{\pi}_{t}\le V_{t}\le\pi_{t}|\boldsymbol{z}^{t-1},\boldsymbol{x}^{t-1},\mathcal{V}^{t-1}(\boldsymbol{d}^{t-1},\boldsymbol{U}^{t-2}(\boldsymbol{d}^{t-2})),\mathcal{U}^{t-1}(\boldsymbol{d}^{t-1},\boldsymbol{y}^{t-1})],
\end{align*}
the sign of which identifies the sign of $\mu_{t}(\boldsymbol{y}^{t-1},\boldsymbol{d}^{t-1},1,x_{t})-\mu_{t}(\boldsymbol{y}^{t-1},\boldsymbol{d}^{t-1},0,\tilde{x}_{t})$
by Assumption RS. The case $t\le2$ can be shown analogously with
$\mathcal{V}^{1}(d_{1})\equiv\mathcal{V}^{1}(d_{1};z_{1})\equiv\{V_{1}:d_{1}=1\{V_{1}\le\pi_{1}(0,0,z_{1})\}\}$.
$\square$

\subsection{Proof of Theorem \ref{thm:ASF}}

\noindent As the first step of identifying $E[Y_{T}(\boldsymbol{d})|\boldsymbol{x},\boldsymbol{z}]$
for given $\boldsymbol{d}=(d_{1},...,d_{T})$, $\boldsymbol{x}=(x_{1},...,x_{T})$
and $\boldsymbol{z}=(z_{1},...,z_{T})$, we apply the result of Lemma
\ref{lem:sign_of_period_ATE}. Fix $t\ge2$ and $\boldsymbol{y}^{t-1}\in\{0,1\}^{t-1}$.
Suppose $x_{t}'$ is such that $\mu_{t}(\boldsymbol{y}^{t-1},\boldsymbol{d}^{t},x_{t})=\mu_{t}(\boldsymbol{y}^{t-1},\boldsymbol{d}^{t-1},d_{t}',x_{t}')$
with $d_{t}'\neq d_{t}$ by applying Lemma \ref{lem:sign_of_period_ATE}.
The existence of $x_{t}'$ is guaranteed by Assumption SP, as $x_{t}\in\mathcal{X}_{t}(\boldsymbol{d}^{t},\boldsymbol{y}^{t-1};\boldsymbol{x}_{-t},\boldsymbol{z}_{-t})\subset\mathcal{X}_{t}(\boldsymbol{d}^{t};\boldsymbol{x}_{-t},\boldsymbol{z}_{-t})$.
Then, as discussed in the main text, $\boldsymbol{U}\in\mathcal{U}(\boldsymbol{d},y_{T};\boldsymbol{x})$
is equivalent to $\boldsymbol{U}\in\mathcal{U}^{T}(d_{t}',\boldsymbol{d}_{-t},y_{T};x_{t}',\boldsymbol{x}_{-t})$
\textit{conditional on} $\boldsymbol{Y}^{t-1}(\boldsymbol{d}^{t-1},\boldsymbol{x}^{t-1})=\boldsymbol{y}^{t-1}$
for all $\boldsymbol{x}_{-t}$ and $\boldsymbol{d}_{-t}$. Then, for
$t\ge2$,
\begin{align}
 & E[Y_{T}(\boldsymbol{d})|\boldsymbol{x},\boldsymbol{z}^{t-1},\boldsymbol{y}^{t-1},\boldsymbol{d}^{t-1},d_{t}']\nonumber \\
= & \Pr\left[\left.\begin{array}{c}
\\
\boldsymbol{U}(\boldsymbol{d})\in\mathcal{U}^{T}(\boldsymbol{d},1;\boldsymbol{x})\\
\\
\end{array}\right|\begin{array}{c}
\boldsymbol{x},\boldsymbol{z}^{t-1},\\
\boldsymbol{U}^{t-1}(\boldsymbol{d}^{t-1})\in\mathcal{U}^{t-1}(\boldsymbol{d}^{t-1},\boldsymbol{y}^{t-1}),\\
\boldsymbol{V}^{t}\in\mathcal{V}^{t}(\boldsymbol{d}^{t-1},d_{t}',\boldsymbol{U}^{t-1}(\boldsymbol{d}^{t-1}))
\end{array}\right]\nonumber \\
= & \Pr\left[\left.\begin{array}{c}
\\
\boldsymbol{U}(\boldsymbol{d})\in\mathcal{U}^{T}(\boldsymbol{d},1;\boldsymbol{x})\\
\\
\end{array}\right|\begin{array}{c}
\boldsymbol{x}_{-t},\boldsymbol{z}^{t},\\
\boldsymbol{U}^{t-1}(\boldsymbol{d}^{t-1})\in\mathcal{U}^{t-1}(\boldsymbol{d}^{t-1},\boldsymbol{y}^{t-1}),\\
\boldsymbol{V}^{t}\in\mathcal{V}^{t}(\boldsymbol{d}^{t-1},d_{t}',\boldsymbol{U}^{t-1}(\boldsymbol{d}^{t-1}))
\end{array}\right],\label{eq:pf_thm1_0}
\end{align}
where the last equality follows from Assumption SX and Lemma \ref{lem:high_level}(ii).
Then, by Assumption RS and the discussion above, \eqref{eq:pf_thm1_0}
is equal to
\begin{align}
 & \Pr\left[\left.\begin{array}{c}
\\
\boldsymbol{U}(d_{t}',\boldsymbol{d}_{-t})\in\mathcal{U}^{T}(d_{t}',\boldsymbol{d}_{-t},1;x_{t}',\boldsymbol{x}_{-t})\\
\\
\end{array}\right|\begin{array}{c}
\boldsymbol{x}_{-t},\boldsymbol{z}^{t},\\
\boldsymbol{U}^{t-1}(\boldsymbol{d}^{t-1})\in\mathcal{U}^{t-1}(\boldsymbol{d}^{t-1},\boldsymbol{y}^{t-1}),\\
\boldsymbol{V}^{t}\in\mathcal{V}^{t}(\boldsymbol{d}^{t-1},d_{t}',\boldsymbol{U}^{t-1}(\boldsymbol{d}^{t-1}))
\end{array}\right]\nonumber \\
= & \Pr\left[\left.\begin{array}{c}
\\
\boldsymbol{U}(d_{t}',\boldsymbol{d}_{-t})\in\mathcal{U}^{T}(d_{t}',\boldsymbol{d}_{-t},1;x_{t}',\boldsymbol{x}_{-t})\\
\\
\end{array}\right|\begin{array}{c}
x_{t}',\boldsymbol{x}_{-t},\boldsymbol{z}^{t},\\
\boldsymbol{U}^{t-1}(\boldsymbol{d}^{t-1})\in\mathcal{U}^{t-1}(\boldsymbol{d}^{t-1},\boldsymbol{y}^{t-1}),\\
\boldsymbol{V}^{t}\in\mathcal{V}^{t}(\boldsymbol{d}^{t-1},d_{t}',\boldsymbol{U}^{t-1}(\boldsymbol{d}^{t-1}))
\end{array}\right]\nonumber \\
= & E[Y_{T}(d_{t}',\boldsymbol{d}_{-t})|x_{t}',\boldsymbol{x}_{-t},\boldsymbol{z}^{t},\boldsymbol{y}^{t-1},\boldsymbol{d}^{t-1},d_{t}'],\label{eq:pf_thm1}
\end{align}
where the first equality is by Assumption SX and Lemma \ref{lem:high_level}.
We use the result \eqref{eq:pf_thm1} in the next step.

First, note that $E[Y_{T}(\boldsymbol{d})|\boldsymbol{x},\boldsymbol{z}^{T},\boldsymbol{y}^{T-1},\boldsymbol{d}^{T}]=E[Y_{T}|\boldsymbol{x},\boldsymbol{z}^{T},\boldsymbol{y}^{T-1},\boldsymbol{d}^{T}]$
is trivially identified for any generic values $(\boldsymbol{d},\boldsymbol{x},\boldsymbol{z},\boldsymbol{y}^{T-1})$.
We prove by means of mathematical induction. For given $2\le t\le T-1$,
suppose $E[Y_{T}(\boldsymbol{d})|\boldsymbol{x},\boldsymbol{z}^{t},\boldsymbol{y}^{t-1},\boldsymbol{d}^{t}]$
is identified for any generic values $(\boldsymbol{d},\boldsymbol{x},\boldsymbol{z}^{t},\boldsymbol{y}^{t-1})$,
and consider the identification of
\begin{align}
E[Y_{T}(\boldsymbol{d})|\boldsymbol{x},\boldsymbol{z}^{t-1},\boldsymbol{y}^{t-2},\boldsymbol{d}^{t-1}]= & \Pr[D_{t}=d_{t}|\boldsymbol{x},\boldsymbol{z}^{t-1},\boldsymbol{y}^{t-2},\boldsymbol{d}^{t-1}]E[Y_{T}(\boldsymbol{d})|\boldsymbol{x},\boldsymbol{z}^{t-1},\boldsymbol{y}^{t-2},\boldsymbol{d}^{t-1},d_{t}]\nonumber \\
 & +\Pr[D_{t}=d_{t}'|\boldsymbol{x},\boldsymbol{z}^{t-1},\boldsymbol{y}^{t-2},\boldsymbol{d}^{t-1}]E[Y_{T}(\boldsymbol{d})|\boldsymbol{x},\boldsymbol{z}^{t-1},\boldsymbol{y}^{t-2},\boldsymbol{d}^{t-1},d_{t}'].\label{eq:expand}
\end{align}
The first main term $E[Y_{T}(\boldsymbol{d})|\boldsymbol{x},\boldsymbol{z}^{t-1},\boldsymbol{y}^{t-2},\boldsymbol{d}^{t-1},d_{t}]$
in \eqref{eq:expand} is identified, by integrating over $y_{t-1}\in\{0,1\}$
the quantity $E[Y_{T}(\boldsymbol{d})|\boldsymbol{x},\boldsymbol{z}^{t-1},\boldsymbol{y}^{t-1},\boldsymbol{d}^{t}]$,
which is assumed to be identified in the previous iteration since
it is equal to $E[Y_{T}(\boldsymbol{d})|\boldsymbol{x},\boldsymbol{z}^{t},\boldsymbol{y}^{t-1},\boldsymbol{d}^{t}]$
by Assumption SX and Lemma \ref{lem:high_level}(ii). The remaining
unknown term in \eqref{eq:expand} satisfies
\begin{align*}
 & E[Y_{T}(\boldsymbol{d})|\boldsymbol{x},\boldsymbol{z}^{t-1},\boldsymbol{y}^{t-2},\boldsymbol{d}^{t-1},d_{t}']\\
= & \Pr[Y_{t-1}=1|\boldsymbol{x},\boldsymbol{z}^{t-1},\boldsymbol{y}^{t-2},\boldsymbol{d}^{t-1},d_{t}']E[Y_{T}(\boldsymbol{d})|\boldsymbol{x},\boldsymbol{z}^{t-1},(\boldsymbol{y}^{t-2},1),\boldsymbol{d}^{t-1},d_{t}']\\
 & +\Pr[Y_{t-1}=0|\boldsymbol{x},\boldsymbol{z}^{t-1},\boldsymbol{y}^{t-2},\boldsymbol{d}^{t-1},d_{t}']E[Y_{T}(\boldsymbol{d})|\boldsymbol{x},\boldsymbol{z}^{t-1},(\boldsymbol{y}^{t-2},0),\boldsymbol{d}^{t-1},d_{t}'].
\end{align*}
By applying \eqref{eq:pf_thm1} to the unknown terms in this expression,
we have
\begin{align}
E[Y_{T}(\boldsymbol{d})|\boldsymbol{x},\boldsymbol{z}^{t-1},\tilde{\boldsymbol{y}}^{t-1},\boldsymbol{d}^{t-1},d_{t}'] & =E[Y_{T}(d_{t}',\boldsymbol{d}_{-t})|x_{t}',\boldsymbol{x}_{-t},\boldsymbol{z}^{t},\tilde{\boldsymbol{y}}^{t-1},\boldsymbol{d}^{t-1},d_{t}']\label{eq:obs_quantity}
\end{align}
for each $\tilde{\boldsymbol{y}}^{t-1}$, which is identified from
the previous iteration. Therefore, $E[Y_{T}(\boldsymbol{d})|\boldsymbol{x},\boldsymbol{z}^{t-1},\boldsymbol{y}^{t-2},\boldsymbol{d}^{t-1}]$
is identified. Note that when $t=2$, $\boldsymbol{Y}^{0}$ is understood
to mean there is no conditioning. Lastly, when $t=1$,
\begin{align*}
E[Y_{T}(\boldsymbol{d})|\boldsymbol{x}]= & \Pr[D_{1}=d_{1}|\boldsymbol{x}]E[Y_{T}(\boldsymbol{d})|\boldsymbol{x},d_{1}]+\Pr[D_{1}=d_{1}'|\boldsymbol{x}]E[Y_{T}(\boldsymbol{d})|\boldsymbol{x},d_{1}'].
\end{align*}
Noting that $Y_{0}=0$, suppose $x_{1}'$ is such that $\mu_{1}(0,d_{1},x_{1})=\mu_{1}(0,d_{1}',x_{1}')$
with $d_{1}'\neq d_{1}$ by applying Lemma \ref{lem:sign_of_period_ATE}.
Then,
\begin{align*}
E[Y_{T}(\boldsymbol{d})|\boldsymbol{x},d_{1}'] & =\Pr[\boldsymbol{U}(\boldsymbol{d})\in\mathcal{U}^{T}(\boldsymbol{d},1;\boldsymbol{x})|\boldsymbol{x}_{-1},z_{1},V_{1}\in\mathcal{V}^{1}(d_{1}')]\\
 & =\Pr[\boldsymbol{U}(d_{1}',\boldsymbol{d}_{-1})\in\mathcal{U}^{T}(d_{1}',\boldsymbol{d}_{-1},1;x_{1}',\boldsymbol{x}_{-1})|\boldsymbol{x}_{-1},z_{1},V_{1}\in\mathcal{V}^{1}(d_{1}')]\\
 & =E[Y_{T}(d_{1}',\boldsymbol{d}_{-1})|x_{1}',\boldsymbol{x}_{-1},z_{1},d_{1}'],
\end{align*}
by Assumption SX and Lemma \ref{lem:high_level}(ii), which is identified
from the previous iteration for $t=2$. Therefore, $E[Y_{T}(\boldsymbol{d})|\boldsymbol{x}]$
is identified. $\Square$

\subsection{Proof of Theorem \ref{thm:TSATE}}

\noindent We analyze the identification of $E[Y_{T}(\boldsymbol{d})|\boldsymbol{Y}_{-}(\boldsymbol{d})=\boldsymbol{y}_{-},\boldsymbol{x},\boldsymbol{z}]$.
Since
\begin{align*}
E[Y_{T}(\boldsymbol{d})|\boldsymbol{Y}_{-}(\boldsymbol{d})=\boldsymbol{y}_{-},\boldsymbol{x},\boldsymbol{z}] & =\Pr[Y_{T}(\boldsymbol{d})=1,\boldsymbol{Y}_{-}(\boldsymbol{d})=\boldsymbol{y}_{-}|\boldsymbol{x},\boldsymbol{z}]/\Pr[\boldsymbol{Y}_{-}(\boldsymbol{d})=\boldsymbol{y}_{-}|\boldsymbol{x},\boldsymbol{z}],
\end{align*}
we identify each term in the fraction. For each term, the proof is
parallel to that of Theorem \ref{thm:ASF}. Let $\tilde{\boldsymbol{y}}_{-}\equiv(y_{1},...,y_{t_{\tilde{L}}})$
be a subvector (not necessarily strict) of $\boldsymbol{y}$, where
$t_{1}<t_{2}<\cdots<t_{\tilde{L}}\le T$ and $\tilde{L}\le T$; e.g.,
when $\tilde{L}=T$, $\tilde{\boldsymbol{y}}_{-}=\boldsymbol{y}$.
Generalizing the $\boldsymbol{U}$-sets introduced in Section \ref{sec:ID},
define
\begin{align*}
\mathcal{U}^{t_{\tilde{L}}}(\boldsymbol{d}^{t_{\tilde{L}}},\tilde{\boldsymbol{y}}_{-})\equiv\mathcal{U}^{t_{\tilde{L}}}(\boldsymbol{d}^{t_{\tilde{L}}},\tilde{\boldsymbol{y}}_{-};\boldsymbol{x}^{t_{\tilde{L}}}) & \equiv\{\boldsymbol{U}^{t_{\tilde{L}}}(\boldsymbol{d}^{t_{\tilde{L}}}):y_{s}=Y_{s}(\boldsymbol{d}^{s},\boldsymbol{x}^{s})\text{ for all }s\in\{t_{1},...,t_{\tilde{L}}\}\}.
\end{align*}
In the first part of the proof, we identify $\Pr[Y_{T}(\boldsymbol{d})=1,\boldsymbol{Y}_{-}(\boldsymbol{d})=\boldsymbol{y}_{-}|\boldsymbol{x},\boldsymbol{z}]$.
Take $\tilde{\boldsymbol{Y}}_{-}(\boldsymbol{d})=(\boldsymbol{Y}_{-}(\boldsymbol{d}),Y_{T}(\boldsymbol{d}))$
with realization $\tilde{\boldsymbol{y}}_{-}=(\boldsymbol{y}_{-},1)$.
For simplicity, we directly use Assumption SX without invoking Lemma
\ref{lem:high_level}(ii). Then, for $2\le t\le T-1$, we have
\begin{align}
 & \Pr[Y_{T}(\boldsymbol{d})=1,\boldsymbol{Y}_{-}(\boldsymbol{d})=\boldsymbol{y}_{-},\boldsymbol{Y}^{t-1}=\boldsymbol{y}^{t-1},\boldsymbol{D}^{t}=(\boldsymbol{d}^{t-1},d_{t}')|\boldsymbol{x},\boldsymbol{z}]\nonumber \\
= & \Pr[\boldsymbol{U}(\boldsymbol{d})\in\mathcal{U}^{T}(\boldsymbol{d},\boldsymbol{y}_{-},1;\boldsymbol{x}),\boldsymbol{U}^{t-1}(\boldsymbol{d}^{t-1})\in\mathcal{U}^{t-1}(\boldsymbol{d}^{t-1},\boldsymbol{y}^{t-1}),\boldsymbol{V}^{t}\in\mathcal{V}^{t}(\boldsymbol{d}^{t-1},d_{t}',\boldsymbol{U}^{t-1}(\boldsymbol{d}^{t-1}))]\nonumber \\
= & \Pr\left[\begin{array}{c}
\boldsymbol{U}(d_{t}',\boldsymbol{d}_{-t})\in\mathcal{U}^{T}(d_{t}',\boldsymbol{d}_{-t},\boldsymbol{y}_{-},1;x_{t}',\boldsymbol{x}_{-t}),\\
\boldsymbol{U}^{t-1}(\boldsymbol{d}^{t-1})\in\mathcal{U}^{t-1}(\boldsymbol{d}^{t-1},\boldsymbol{y}^{t-1}),\\
\boldsymbol{V}^{t}\in\mathcal{V}^{t}(\boldsymbol{d}^{t-1},d_{t}',\boldsymbol{U}^{t-1}(\boldsymbol{d}^{t-1}))
\end{array}\right]\nonumber \\
= & \Pr[Y_{T}(d_{t}',\boldsymbol{d}_{-t})=1,\boldsymbol{Y}_{-}(d_{t}',\boldsymbol{d}_{-t})=\boldsymbol{y}_{-},\boldsymbol{Y}^{t-1}=\boldsymbol{y}^{t-1},\boldsymbol{D}^{t}=(\boldsymbol{d}^{t-1},d_{t}')|x_{t}',\boldsymbol{x}_{-t},\boldsymbol{z}],\label{eq:pf_thm1-1}
\end{align}
where the second equality uses $x_{t}'$ such that $\mu_{t}(\boldsymbol{y}^{t-1},\boldsymbol{d}^{t},x_{t})=\mu_{t}(\boldsymbol{y}^{t-1},\boldsymbol{d}^{t-1},d_{t}',x_{t}')$
by applying Lemma \ref{lem:sign_of_period_ATE}. First, $\Pr[Y_{T}(\boldsymbol{d})=1,\boldsymbol{Y}_{-}(\boldsymbol{d})=\boldsymbol{y}_{-},\boldsymbol{Y}^{T-1}=\boldsymbol{y}^{T-1},\boldsymbol{D}^{T}=\boldsymbol{d}^{T}|\boldsymbol{x},\boldsymbol{z}]=\Pr[Y_{T}=1,\boldsymbol{Y}_{-}=\boldsymbol{y}_{-},\boldsymbol{Y}^{T-1}=\boldsymbol{y}^{T-1},\boldsymbol{D}^{T}=\boldsymbol{d}^{T}|\boldsymbol{x},\boldsymbol{z}]$
is trivially identified for any generic values $(\boldsymbol{d},\boldsymbol{x},\boldsymbol{z},\boldsymbol{y}^{T-1})$.
For given $2\le t\le T-1$, suppose $\Pr[Y_{T}(\boldsymbol{d})=1,\boldsymbol{Y}_{-}(\boldsymbol{d})=\boldsymbol{y}_{-},\boldsymbol{Y}^{t-1}=\boldsymbol{y}^{t-1},\boldsymbol{D}^{t}=\boldsymbol{d}^{t}|\boldsymbol{x},\boldsymbol{z}]$
is identified for any generic values $(\boldsymbol{d},\boldsymbol{x},\boldsymbol{z},\boldsymbol{y}^{t-1})$,
and consider identification of
\begin{align}
 & \Pr[Y_{T}(\boldsymbol{d})=1,\boldsymbol{Y}_{-}(\boldsymbol{d})=\boldsymbol{y}_{-},\boldsymbol{Y}^{t-2}=\boldsymbol{y}^{t-2},\boldsymbol{D}^{t-1}=\boldsymbol{d}^{t-1}|\boldsymbol{x},\boldsymbol{z}]\nonumber \\
= & \Pr[Y_{T}(\boldsymbol{d})=1,\boldsymbol{Y}_{-}(\boldsymbol{d})=\boldsymbol{y}_{-},\boldsymbol{Y}^{t-2}=\boldsymbol{y}^{t-2},\boldsymbol{D}^{t}=(\boldsymbol{d}^{t-1},d_{t})|\boldsymbol{x},\boldsymbol{z}]\nonumber \\
 & +\Pr[Y_{T}(\boldsymbol{d})=1,\boldsymbol{Y}_{-}(\boldsymbol{d})=\boldsymbol{y}_{-},\boldsymbol{Y}^{t-2}=\boldsymbol{y}^{t-2},\boldsymbol{D}^{t}=(\boldsymbol{d}^{t-1},d_{t}')|\boldsymbol{x},\boldsymbol{z}].\label{eq:unobs_exp}
\end{align}
The first term in the expression is identified, by summing over $y_{t-1}$
the quantity $\Pr[Y_{T}(\boldsymbol{d})=1,\boldsymbol{Y}_{-}(\boldsymbol{d})=\boldsymbol{y}_{-},\boldsymbol{Y}^{t-1}=\boldsymbol{y}^{t-1},\boldsymbol{D}^{t}=\boldsymbol{d}^{t}|\boldsymbol{x},\boldsymbol{z}]$,
which is identified from the previous iteration. The second unknown
term in \eqref{eq:unobs_exp} satisfies
\begin{align}
 & \Pr[Y_{T}(\boldsymbol{d})=1,\boldsymbol{Y}_{-}(\boldsymbol{d})=\boldsymbol{y}_{-},\boldsymbol{Y}^{t-2}=\boldsymbol{y}^{t-2},\boldsymbol{D}^{t}=(\boldsymbol{d}^{t-1},d_{t}')|\boldsymbol{x},\boldsymbol{z}]\nonumber \\
= & \Pr[Y_{T}(\boldsymbol{d})=1,\boldsymbol{Y}_{-}(\boldsymbol{d})=\boldsymbol{y}_{-},\boldsymbol{Y}^{t-1}=(\boldsymbol{y}^{t-2},1),\boldsymbol{D}^{t}=(\boldsymbol{d}^{t-1},d_{t}')|\boldsymbol{x},\boldsymbol{z}]\nonumber \\
 & +\Pr[Y_{T}(\boldsymbol{d})=1,\boldsymbol{Y}_{-}(\boldsymbol{d})=\boldsymbol{y}_{-},\boldsymbol{Y}^{t-1}=(\boldsymbol{y}^{t-2},0),\boldsymbol{D}^{t}=(\boldsymbol{d}^{t-1},d_{t}')|\boldsymbol{x},\boldsymbol{z}].\label{eq:unobs_expansion-1}
\end{align}
But note that, by \eqref{eq:pf_thm1-1}, each term in \eqref{eq:unobs_expansion-1}
satisfies
\begin{align}
 & \Pr[Y_{T}(\boldsymbol{d})=1,\boldsymbol{Y}_{-}(\boldsymbol{d})=\boldsymbol{y}_{-},\boldsymbol{Y}^{t-1}=\tilde{\boldsymbol{y}}^{t-1},\boldsymbol{D}^{t}=(\boldsymbol{d}^{t-1},d_{t}')|\boldsymbol{x},\boldsymbol{z}]\nonumber \\
= & \Pr[Y_{T}(d_{t}',\boldsymbol{d}_{-t})=1,\boldsymbol{Y}_{-}(d_{t}',\boldsymbol{d}_{-t})=\boldsymbol{y}_{-},\boldsymbol{Y}^{t-1}=\tilde{\boldsymbol{y}}^{t-1},\boldsymbol{D}^{t}=(\boldsymbol{d}^{t-1},d_{t}')|x_{t}',\boldsymbol{x}_{-t},\boldsymbol{z}]\label{eq:unobs_expansion2-1}
\end{align}
for each $\tilde{\boldsymbol{y}}^{t-1}$, which is identified from
the previous iteration. Therefore, $\Pr[Y_{T}(\boldsymbol{d})=1,\boldsymbol{Y}_{-}(\boldsymbol{d})=\boldsymbol{y}_{-},\boldsymbol{Y}^{t-2}=\boldsymbol{y}^{t-2},\boldsymbol{D}^{t-1}=\boldsymbol{d}^{t-1}|\boldsymbol{x},\boldsymbol{z}]$
is identified. Lastly, when $t=1$,
\begin{align*}
\Pr[Y_{T}(\boldsymbol{d})=1,\boldsymbol{Y}_{-}(\boldsymbol{d})=\boldsymbol{y}_{-}|\boldsymbol{x},\boldsymbol{z}]= & \Pr[Y_{T}(\boldsymbol{d})=1,\boldsymbol{Y}_{-}(\boldsymbol{d})=\boldsymbol{y}_{-},D_{1}=d_{1}|\boldsymbol{x},\boldsymbol{z}]\\
 & +\Pr[Y_{T}(\boldsymbol{d})=1,\boldsymbol{Y}_{-}(\boldsymbol{d})=\boldsymbol{y}_{-},D_{1}=d_{1}'|\boldsymbol{x},\boldsymbol{z}].
\end{align*}
The first term is identified from the iteration for $t=2$. Noting
that $Y_{0}=0$, suppose $x_{1}'$ is such that $\mu_{1}(0,d_{1},x_{1})=\mu_{1}(0,d_{1}',x_{1}')$
with $d_{1}'\neq d_{1}$ by Lemma \ref{lem:sign_of_period_ATE}. Then,
similarly to \eqref{eq:pf_thm1-1},
\begin{align*}
 & \Pr[Y_{T}(\boldsymbol{d})=1,\boldsymbol{Y}_{-}(\boldsymbol{d})=\boldsymbol{y}_{-},D_{1}=d_{1}'|\boldsymbol{x},\boldsymbol{z}]\\
= & \Pr[Y_{T}(d_{1}',\boldsymbol{d}_{-1})=1,\boldsymbol{Y}_{-}(d_{1}',\boldsymbol{d}_{-1})=\boldsymbol{y}_{-},D_{1}=d_{1}'|x_{1}',\boldsymbol{x}_{-1},\boldsymbol{z}],
\end{align*}
which is also identified from the previous iteration for $t=2$. Therefore
$\Pr[Y_{T}(\boldsymbol{d})=1,\boldsymbol{Y}_{-}(\boldsymbol{d})=\boldsymbol{y}_{-}|\boldsymbol{x},\boldsymbol{z}]$
is identified.

In the second part of the proof, we identify $\Pr[\boldsymbol{Y}_{-}(\boldsymbol{d})=\boldsymbol{y}_{-}|\boldsymbol{x},\boldsymbol{z}]$.
Take $\tilde{\boldsymbol{Y}}_{-}(\boldsymbol{d})=\boldsymbol{Y}_{-}(\boldsymbol{d})\equiv(Y_{t_{1}}(\boldsymbol{d}),...,Y_{t_{L}}(\boldsymbol{d}))$
with realization $\tilde{\boldsymbol{y}}_{-}=\boldsymbol{y}_{-}$.
Then, for $2\le t\le t_{L}-1$, we can show the following equivalence,
analogous to \eqref{eq:pf_thm1-1}:
\begin{align*}
 & \Pr[\boldsymbol{Y}_{-}(\boldsymbol{d})=\boldsymbol{y}_{-},\boldsymbol{Y}^{t-1}=\boldsymbol{y}^{t-1},\boldsymbol{D}^{t}=(\boldsymbol{d}^{t-1},d_{t}')|\boldsymbol{x},\boldsymbol{z}]\\
= & \Pr[\boldsymbol{U}^{t_{L}}(\boldsymbol{d}^{t_{L}})\in\mathcal{U}^{t_{L}}(\boldsymbol{d}^{t_{L}},\boldsymbol{y}_{-};\boldsymbol{x}^{t_{L}}),\boldsymbol{U}^{t-1}(\boldsymbol{d}^{t-1})\in\mathcal{U}^{t-1}(\boldsymbol{d}^{t-1},\boldsymbol{y}^{t-1}),\boldsymbol{V}^{t}\in\mathcal{V}^{t}(\boldsymbol{d}^{t-1},d_{t}',\boldsymbol{U}^{t-1}(\boldsymbol{d}^{t-1}))]\\
= & \Pr\left[\begin{array}{c}
\boldsymbol{U}^{t_{L}}(d_{t}',\boldsymbol{d}_{-t}^{t_{L}})\in\mathcal{U}^{t_{L}}(d_{t}',\boldsymbol{d}_{-t}^{t_{L}},\boldsymbol{y}_{-};x_{t}',\boldsymbol{x}_{-t}^{t_{L}}),\\
\boldsymbol{U}^{t-1}(\boldsymbol{d}^{t-1})\in\mathcal{U}^{t-1}(\boldsymbol{d}^{t-1},\boldsymbol{y}^{t-1}),\\
\boldsymbol{V}^{t}\in\mathcal{V}^{t}(\boldsymbol{d}^{t-1},d_{t}',\boldsymbol{U}^{t-1}(\boldsymbol{d}^{t-1}))
\end{array}\right]\\
= & \Pr[\boldsymbol{Y}_{-}(d_{t}',\boldsymbol{d}_{-t})=\boldsymbol{y}_{-},\boldsymbol{Y}^{t-1}=\boldsymbol{y}^{t-1},\boldsymbol{D}^{t}=(\boldsymbol{d}^{t-1},d_{t}')|x_{t}',\boldsymbol{x}_{-t},\boldsymbol{z}].
\end{align*}
The rest of the proof is an immediate modification of the iterative
argument in the first part, and hence is omitted. $\square$

\end{appendix}
\end{document}